\documentclass{aa}  

\usepackage{graphicx}
\usepackage{txfonts}
\usepackage{natbib}
\usepackage{xcolor}
\usepackage[normalem]{ulem}

\definecolor{linkblue}{rgb}{0.0,0.0,0.45}
\usepackage[colorlinks=true,
  linkcolor=blue,
  citecolor=linkblue,
  urlcolor=blue]{hyperref}
  
\usepackage{placeins}           

\newcommand{\pker}[1]{U^{({\textsc{#1})}}}
\newcommand{\pspc}[1]{f^{({\textsc{#1})}}}

\newcommand{\tpker}[1]{\tilde{U}^{({\textsc{#1})}}}

\begin{document}

   \title{Faint Companions to  Hot Stars}

   \subtitle{Spectroscopic Detection Sensitivity and Implications for Survey Design}

%
%
%

   \author{Sahar Shahaf\inst{1}\corrauth{sashahaf@mpia.de}        
        \and{{Hans-Walter Rix}\inst{1}}
        \and{{Johanna M\"uller-Horn}\inst{1}}
        \and{{Jaime~I. Villase\~nor}\inst{1}}
        \and{{Tomer Shenar}\inst{2}}
        \and{{Rhys Seeburger}\inst{3}}
        }

   \institute{Max-Planck-Institut f\"ur Astronomie (MPIA), K\"onigstuhl 17, 69117 Heidelberg, Germany.
   \and
   Tel Aviv University, The School of Physics and Astronomy, Tel Aviv 6997801, Israel.
   \and
   Astrophysics Research Institute, Liverpool John Moores University, 146 Brownlow Hill, Liverpool L3 5RF, UK.}

   \date{Accepted XXX. Received YYY}


\abstract{
Searches for black holes in binaries must distinguish genuinely dark companions from faint luminous stars. This is particularly challenging for massive hot stars. Their masses can exceed those of typical stellar black holes, while their spectra contain only a few broad lines.
We develop a Fourier-domain method that converts isolated, broadened spectral lines into calibrated constraints on faint companions. In Fourier space, Doppler shifts are encoded in the phase, while the line profile defines a smooth amplitude envelope. A faint secondary introduces an oscillatory perturbation to the otherwise smooth log-amplitude. Expanding this perturbation to first order in the secondary-to-primary line-depth ratio, $\alpha$, allows it to be profiled against the smooth baseline in a likelihood-ratio test with the single-star model.
Because the model is analytic and nearly linear, its sensitivity can be evaluated transparently for individual spectra and at survey scale. It can be applied to single or multiple epochs without requiring full spectral disentangling.
At small velocity separation $v$, the companion signal becomes degenerate with the smooth single-star profile. For the Fourier log-amplitude, the fractional uncertainty scales as $\Delta\alpha/\alpha \propto v^{-4}$. If the primary velocity is known, the Fourier phase provides a more favorable detection channel whose sensitivity scales as $v^{-1}$.
We demonstrate the method on archival FEROS spectra of the massive binary CPD$-41^\circ$7742. We detect the known secondary in He\,I~$6678$~\AA\ and obtain a stringent non-detection in He\,II~$5411$~\AA, where its contribution is expected to be negligible.
The same framework yields local detection efficiencies as functions of line-depth ratio, velocity separation, noise, and instrumental setup. These can be combined across lines, epochs, and targets to construct survey selection functions for spectroscopic black-hole searches, including those enabled by SDSS-V and Gaia.
}


\maketitle
\nolinenumbers

\section{Introduction}

Massive stars (${\gtrsim}8\,M_\odot$) are the progenitors of neutron stars and black holes. Many evolve in binary systems, through pathways that shape the properties of compact-object remnants \citep{Marchant2024}. Massive-star binaries with black-hole companions offer a direct probe of these pathways and have motivated searches for quiescent black holes. More broadly, detached post-interaction binaries encode which systems survive these evolutionary phases and constrain natal kicks, fallback retention, and envelope ejection \citep[see, for example,][]{Fryer2012,Giacobbo2018,Willcox2025}. Large, homogeneous surveys now make it possible to study these populations at scale.

Among massive stars, the fraction of systems with black-hole companions is expected to be of order a few percent \citep{Langer2020}. The identification of these systems, however, is intrinsically challenging. Less massive luminous companions can reproduce similar dynamical signatures while contributing only weakly to the observed spectrum \citep[e.g.,][]{Shenar2020,Bodensteiner2020}. Distinguishing these luminous companions from black holes therefore requires either detecting their spectral contribution or establishing that such a contribution would have been detectable.

Early-type spectra contain few diagnostic lines, which are often broadened by rapid rotation, with projected rotational velocities of ${\sim}100$~$\mathrm{km\,s^{-1}}$ or more. In this regime, spectroscopic techniques lose sensitivity and blends become difficult to interpret \citep{Bodensteiner2020,Muller-Horn2026,Rowan2026}. The inferred properties of the luminous star may also be biased, leading to incorrect conclusions about the nature of the companion \citep[e.g.,][]{El-Badry2021}; a well-known example is LB-1, where an apparent companion signature was later traced to the broad line profile of the primary \citep{Liu2019,Abdul-Masih2020}.

Several established methods address related aspects of this problem. Two-dimensional cross-correlation techniques, such as \texttt{TODCOR}, search for binary signatures using two templates simultaneously \citep{Zucker1994}, while spectral disentangling and secondary-template reconstruction methods recover component spectra from multi-epoch data \citep[e.g.,][]{Gonzalez2006,Gonzalez2024,Seeburger2024}. More recently, data-driven spectral models and deep-learning approaches have been used to identify faint companions in large spectroscopic data sets \citep{Ness2015,Angelo2026,Binnenfeld2025}. Beyond detection, these techniques can place upper limits on the luminosity of an unseen companion, which has been central to the identification and validation of quiescent black-hole companions to massive stars \citep{Shenar2022,Mahy2022, Bodensteiner2022}.

Rotational broadening, line depth, and velocity separation all affect the ability to identify, or exclude, a secondary contribution to the observed light. More generally, analysis techniques can lose sensitivity in ways that are difficult to quantify a priori. In searches for rare non-luminous companions, this limitation translates into characterization ambiguity, which can sometimes be resolved with complementary observations \citep[e.g.,][]{Rowan2025,Rowan2026}. In large surveys, the same limitation becomes a selection-function problem \citep{Rix2021,Essick2026}. In both cases, a non-detection of secondary light is informative only if the detection efficiency can be quantified with sufficient fidelity.

Population studies and individual-system characterization therefore require practical ways to determine which companions would have been detected, and with what efficiency. Massive-star spectra offer a useful simplification. When the relevant information is carried by isolated, broadened lines, the detailed stellar spectrum can often be replaced by a low-dimensional approximation. This makes it possible to build search schemes with minimal assumptions about the underlying spectrum and to quantify the sensitivity of a given survey. 

To address this need, we develop a Fourier-domain detection framework for isolated, broadened spectral lines. Fourier analysis of stellar spectra has a long history \citep[e.g.,][and references therein]{Gray1973,Smith1976,Smith1976b, Gray1992}. Here, we use it to quantify when a faint secondary produces a detectable distortion of the primary line profile. We derive analytic detection limits in the unresolved regime and demonstrate the approach on archival spectra of CPD$-$41$^\circ$~7742. The resulting line-by-line limits provide the local detection efficiencies needed to interpret non-detections and identify luminous impostors in spectroscopic searches for black holes.


\section{Spectroscopy of isolated and composite lines}\label{sec: spectroscopy}
We develop the approach in two stages. We first treat an isolated line from a single star; its Fourier log-amplitude forms a smooth envelope. This provides a simple rotational-velocity measurement and illustrates the central principle of the method: only structure that cannot be absorbed by the fitted line profile remains measurable. In subsequent sections, we add a faint companion and show that its signature appears in the same residual as an orbital-phase-dependent perturbation.

\subsection{A single star}\label{sec: isolated lines}
We consider the limit in which the spectral lines are isolated and external broadening dominates over the intrinsic line profile. Following the notation of \citet{Shahaf2023, Shahaf2026}, we denote the local rest-frame spectrum of star $\textsc{A}$ by $\pspc{a}$. Its convolution with the principal kernel, $\pker{a}$, yields the disk-integrated profile. The kernel aggregates the weighted contributions of all visible surface elements and accounts for rotation, thermal broadening, pressure broadening, and instrumental effects. The label $A$ identifies the star, not a surface element.
Following \citet{Tonry1979}, we resample the spectra onto a logarithmic wavelength grid and express its coordinate, $\lambda$, in velocity units. A radial-velocity shift $\delta v$ then corresponds to a translation $\delta\lambda=c\ln(1+\delta v/c)\simeq\delta v$ along this coordinate.

The projected-area-weighted integration over the visible stellar surface is
encoded in the principal kernel, $\pker{a}$. For pure rotational broadening, for example, the kernel represents the distribution of local line-of-sight Doppler shifts, weighted by the limb-darkened flux contributed by each surface element.\footnote{See the discussion below and \citet{Shahaf2023} for the formal derivation of the principal kernel. For convenience in developing the equations here, we depart slightly from the convention adopted there by absorbing the Doppler shift into the line profile $\pspc{a}$ itself.} The resulting disk-integrated, continuum-normalized, baseline-subtracted spectrum of a single star is
\begin{equation}\label{eq: sb1 convolution}
S(\lambda) = \pker{a}(\lambda) \circledast \pspc{a}(\lambda). 
\end{equation}

When the intrinsic line profile is narrow compared to the external broadening scales, its detailed structure becomes negligible and the line can be effectively treated as a delta function. Therefore, %
\begin{equation}
\pspc{a}(\lambda) \simeq -\alpha_{\textsc{a}} \delta\left[\lambda - (\lambda_0+ v_{\textsc{a}})\right],
\end{equation}
where $\lambda_0$ is the rest-frame line center, $v_\textsc{a}$ is the center-of-mass radial velocity, and $\alpha_a$ sets the line normalization. We assume that the local line strength is uniform across the stellar disk; the projected-area and limb-darkening weights enter the principal kernel.
The Fourier transform converts the convolution in equation~(\ref{eq: sb1 convolution}) into a product. In the delta-function limit, 
\begin{equation} \label{eq: sb1 Stilde}
\tilde{S}(\zeta) \simeq -\alpha_{\textsc{a}} \,
{\tpker{a}}(\zeta)\,
\exp\left[-i2\pi \zeta (\lambda_0 + v_{\textsc{a}})\right],
\end{equation}
where $\zeta$ is the conjugate-wavelength coordinate with units reciprocal to velocity. 

\begin{figure}
    \includegraphics[width=0.9\linewidth]{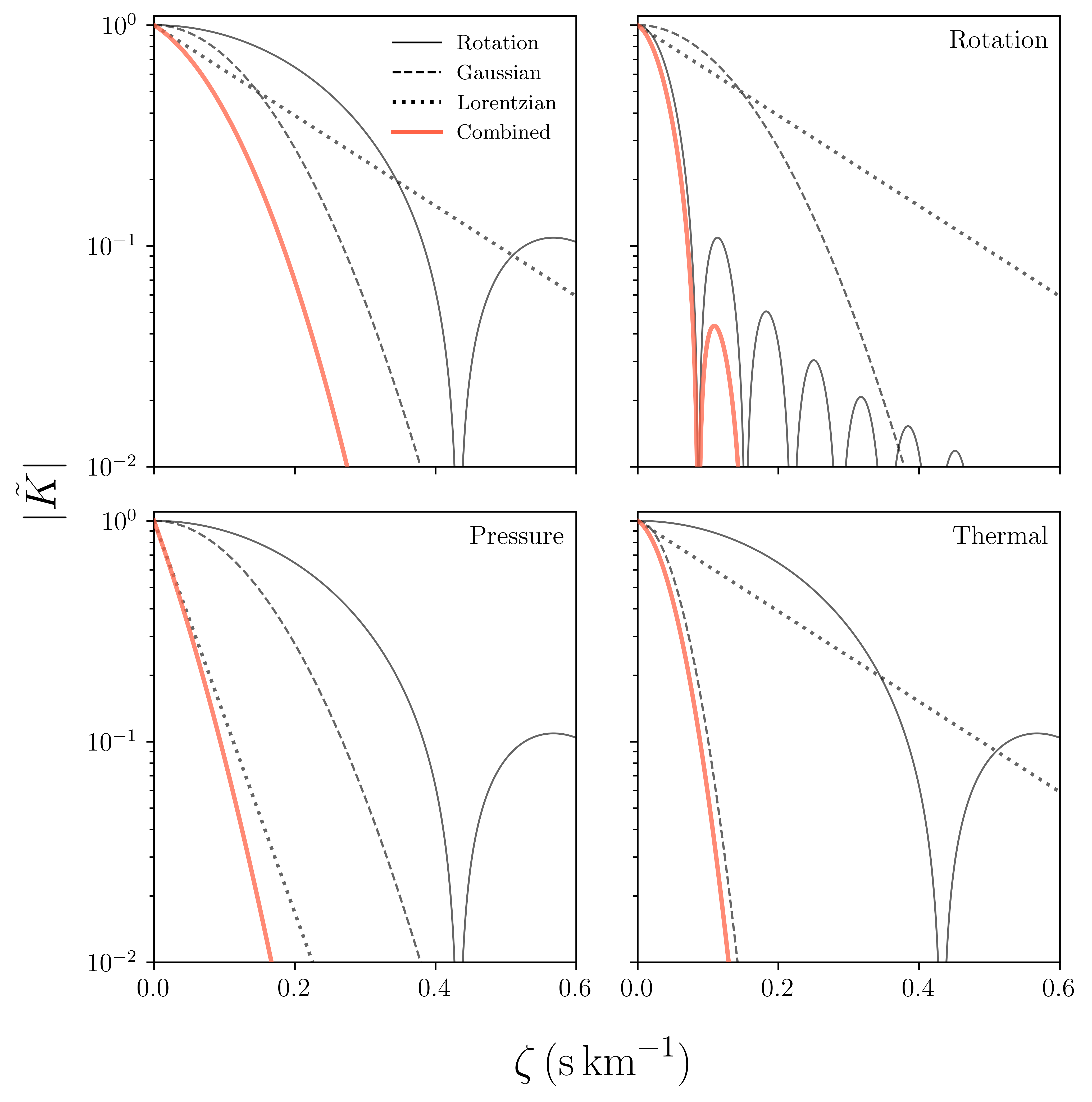}
\caption{Fourier-domain amplitudes of the broadening kernels. The rotational, Gaussian, and Lorentzian terms are shown as thin solid, dashed, and dotted black curves, respectively. Their product, defined in equation~(\ref{eq: kernel terms}), is shown as a thick red curve. The top-left panel shows a case in which the three broadening components contribute comparably. The remaining panels show rotation-dominated, thermal-dominated, and pressure-dominated cases, as labeled.}
    \label{fig: kernel example}
\end{figure}

\begin{figure*}
    \centering
    \includegraphics[width=1\linewidth]{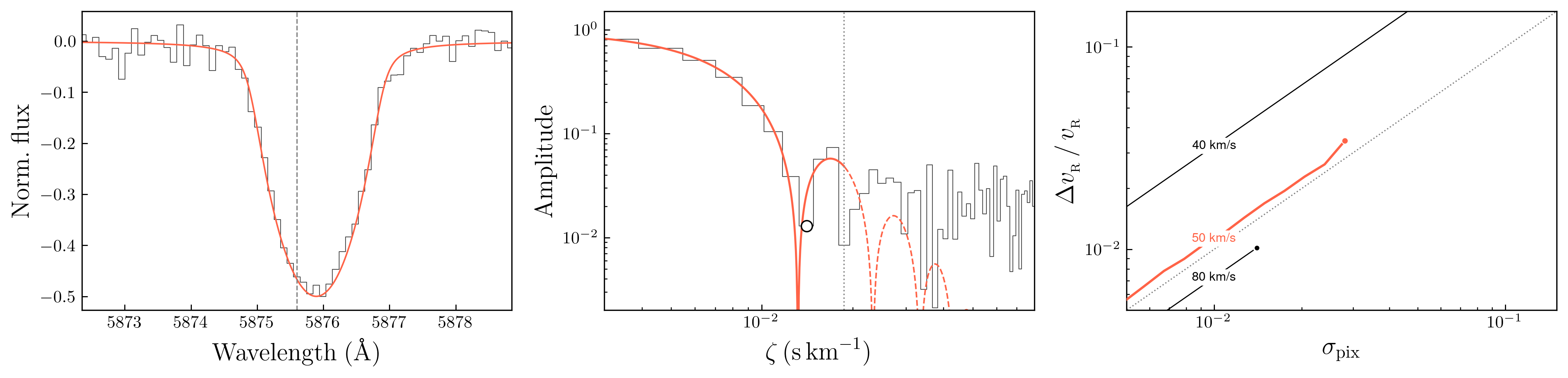}
    \caption{Simulated recovery of $v\sin i$ from a single line in the Fourier domain. Left: Rotationally broadened absorption line with $v_{\textsc{r}}=50$, $v_{\textsc{g}}=10$, $v_\gamma=10~\mathrm{km\,s^{-1}}$, and $\epsilon=0.5$. The high-resolution profile is shown in red; the black step curve shows sampling at $5~\mathrm{km\,s^{-1}}$ with Gaussian noise $\sigma_{\rm pix}=0.02$. The vertical dashed line marks the nominal line center. Middle: Normalized Fourier amplitude of the noisy spectrum and fitted broadening kernel. The fit gives $v_{\textsc{r}}\simeq49~\mathrm{km\,s^{-1}}$. Only modes below the fixed limit $\zeta\le0.02~\mathrm{s\,km^{-1}}$ (dotted line) are fitted; this range includes the first zero and the first sidelobe of the rotational kernel, where the sensitivity to $v_{\textsc{r}}$ resides (Appendix~\ref{app: expected precision}). The model is dashed beyond the limit. The first log-amplitude minimum is marked by an open circle. Choosing this position as the first zero of the Bessel function in equation~(\ref{eq: rotational kernel}) would imply $v_{\textsc{r}}\simeq46~\mathrm{km\,s^{-1}}$ using the correct limb-darkening coefficient. {Right:} Fractional precision $\Delta v_{\textsc{r}}/v_{\textsc{r}}$ as a function of per-pixel noise. Black curves show the Cram\'er--Rao bounds for $v\sin i=40$ and $80~\mathrm{km\,s^{-1}}$ lines, at fixed equivalent width, and the red curve shows an empirical estimate for $v_{\textsc{r}}=50~\mathrm{km\,s^{-1}}$. Curves terminate where fewer than $7$ Fourier modes of the noise-free model exceed a signal-to-noise ratio of $3$. The gray dotted line marks a one-to-one relation for reference. }
    \label{fig: line example}
\end{figure*}

We decompose the principal kernel into Lorentzian, Gaussian, and rotational components, characterized by velocity scales $v_\gamma$, $v_{\textsc{g}}$, and $v_{\textsc{r}}$, respectively. The first two are the full widths at half maximum (FWHM) of the corresponding profiles in velocity space, while $v_{\textsc{r}}$ is the projected rotational velocity. The Lorentzian term represents pressure broadening, the Gaussian term includes thermal and instrumental broadening, and the rotational term accounts for disk-integrated rotation.
As shown by \citet{Gray1992}, the Fourier transform of the kernel can be expressed as
\begin{equation} \label{eq: kernel terms}
\tpker{a}(\zeta)
=
\tilde{K}_{\gamma}(\zeta\,;\,v_\gamma)\,
\tilde{K}_{\textsc{g}}(\zeta\,;\,v_{\textsc{g}})\,
\tilde{K}_{\textsc{r}}(\zeta\,;\,v_{\textsc{r}}, \epsilon),
\end{equation}
where $\epsilon$ is the linear limb-darkening coefficient.  For hot stars, macroturbulent broadening is often comparable to or larger than rotation \citep{Simon-Diaz2014}; here it is subsumed into the effective Gaussian term, which is adequate as long as its detailed profile is not resolved.

Equation~\eqref{eq: sb1 Stilde} approximates the intrinsic line profile as a shifted delta function. Its displacement appears as a complex phase in the Fourier domain. Taking the Fourier amplitude removes this phase, and hence the Doppler-shift dependence, yielding
$|\tilde{S}(\zeta)| \simeq \alpha_{\textsc{a}}|\tpker{a}(\zeta)|$. In this approximate description, the amplitude fully describes the line shape. Its logarithm,
\begin{equation}\label{eq: lnS}
\ln |\tilde{S}| \simeq
\ln |\tpker{a}| + \mathrm{const},
\end{equation}
converts the product of broadening kernels into a sum and makes the model linear in the non-rotational broadening terms. The kernel functions are given explicitly in Appendix~\ref{app: kernels}, and their contributions to the log-amplitude are illustrated in Figure~\ref{fig: kernel example}. The resulting model is
\begin{equation}\label{eq: lnS components}
\ln |\tilde{S}(\zeta)| \simeq
-{\pi v_\gamma} |\zeta|
-
\frac{\pi^2 v_{\textsc{g}}^2}{4\ln2} \zeta^2
+
k(\zeta; v_{\textsc{r}}, \epsilon)
+
\mathrm{const},
\end{equation}
where, for convenience, we denote the logarithm of the rotational contribution as $k$. This representation motivates a simple fitting scheme for the measured logarithmic Fourier amplitudes. Since the data are real-valued, we consider only non-negative $\zeta$ values, and model the Fourier log-amplitudes as
\begin{equation}\label{eq: single line model}
y_i = a_0 + a_1\zeta_i + a_2\zeta_i^2
      + k(\zeta_i; a_3,\epsilon) + {\rm noise},
\end{equation}
where $y_i  \equiv \ln{ |\tilde{S}(\zeta_i)|}$ and the coefficients $(a_0,a_1,a_2)$ describe the smooth non-rotational envelope, while $a_3$ sets the projected rotational velocity.  For fixed $a_3$, the remaining parameters are linear and can be profiled out. The sensitivity to rotation is therefore set by the component of $k(\zeta_i;a_3,\epsilon)$ that remains in the residuals after fitting the polynomial baseline. The additive noise term is a linearized approximation, whose scaling and validity are discussed in Appendix~\ref{app: noise model}.

We now compare estimates of the projected rotational velocity obtained by locating the first Fourier minimum and by fitting the log-amplitude profile. The left and middle panels of Figure~\ref{fig: line example} show a rotationally broadened spectral line and its corresponding Fourier log-amplitude spectrum. This representation builds on the classical Fourier treatment of rotational broadening \citep[e.g.,][]{Gray1992,Simon-Diaz2007,Simon-Diaz2014}. These studies often infer the projected rotational velocity from the first zero of the rotational kernel, visible as a notch in the middle panel. Its location follows from equation~(\ref{eq: rotational kernel}) and scales inversely with the rotational velocity, although additional broadening and noise can obscure it. The first minimum in the middle panel is marked by an open circle. Its position implies $v_{\textsc{r}}\simeq46~\mathrm{km\,s^{-1}}$ for the adopted limb-darkening coefficient and $v_{\textsc{r}}\simeq43~\mathrm{km\,s^{-1}}$ for $\epsilon=0$, compared with the injected value of $50~\mathrm{km\,s^{-1}}$. We estimate the uncertainty associated with the finite Fourier grid by assuming that the continuous position of the minimum is uniformly distributed within one grid interval. This gives a formal sampling uncertainty of $1.5~\mathrm{km\,s^{-1}}$ for the present example (see an alternative treatment of sampling uncertainty by \citealt{Royer2002}).

\begin{figure*}
    \centering
    \includegraphics[width=1\linewidth]{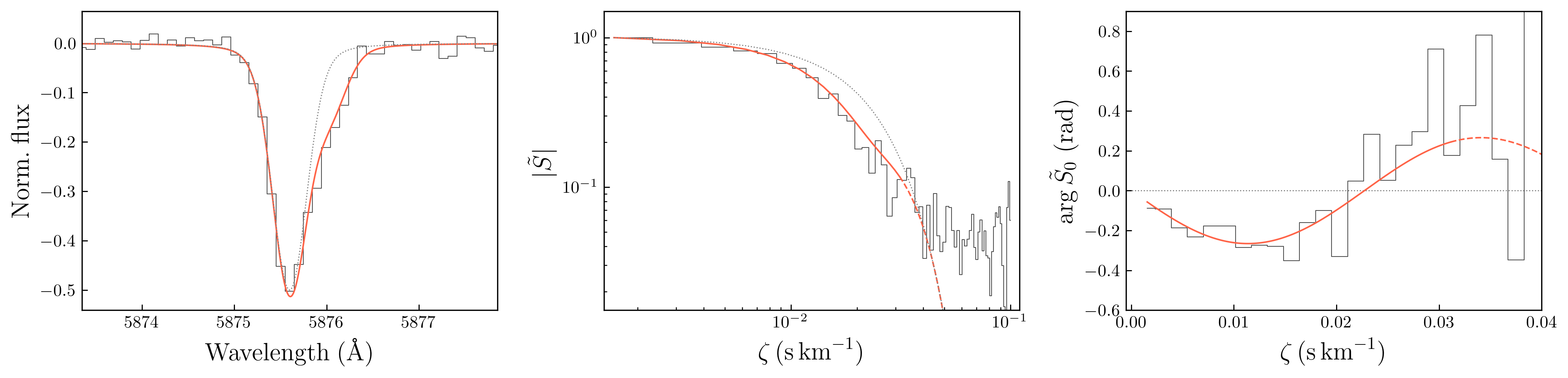}
\caption{
Recovery of the line-depth ratio in the line-spread-function-limited regime. 
Left: simulated binary composite spectrum of two components separated by $22\,\mathrm{km\,s^{-1}}$, $\alpha=0.25$, and per-pixel noise $\sigma_{\rm pix}=0.015$; the dotted curve shows the primary alone. 
Middle: normalized Fourier amplitude of the same realization, showing the fitted oscillatory binary perturbation (solid red line) to the single-star reference envelope (dotted black line). The dashed red line is an extrapolated model, overlaid beyond the limit of the fitted domain.
Right: the argument of the shifted Fourier transformed spectrum, and the corresponding fitted model, plotted according to the conventions of the middle panel. 
}
    \label{fig: double line example}
\end{figure*}

This estimate is conditional on correctly identifying the minimum. At the adopted noise level, 500 independent realizations yield an average minimum-based velocity consistent with the injected value, with a scatter of $3.1~\mathrm{km\,s^{-1}}$. The offset in the displayed realization is therefore consistent with the expected scatter rather than a systematic bias. Repeating the calculation over a range of noise levels shows that the scatter increases with noise. Once spurious minima at lower Fourier frequencies become common, the minimum-based estimate becomes biased towards higher velocities. By comparison, the profiled $\chi^2$ fit to the Fourier-amplitude profile of the displayed realization gives $v_{\textsc{r}}=49.3~\mathrm{km\,s^{-1}}$ with a formal uncertainty of $0.9~\mathrm{km\,s^{-1}}$. Injection--recovery over 3000 independent noise realizations shows that this formal 1$\sigma$ confidence interval covers ${\sim}\,60\%$ of the recovered values, implying that the formal uncertainty mildly underestimates the spread. Scaling the formal error to the nominal $68\%$ coverage gives a realistic uncertainty estimate of $1.1~\mathrm{km\,s^{-1}}$ (see Appendix~\ref{app: noise model}).

In the simulated example, the fit uses a fixed band of $\zeta\le0.02~\mathrm{s\,km^{-1}}$ (dotted line in the middle panel of Figure~\ref{fig: line example}). The band includes the first zero and the first sidelobe of the rotational kernel, beyond which the noise-free amplitude falls below the noise level (Appendix~\ref{app: noise model}). At low $\zeta$, the rotational contribution is well approximated by a fourth-order polynomial. Its distinctive structure appears at higher frequencies, around the first zero of the Bessel function. 
Precise measurements of rotational broadening therefore require sufficient signal-to-noise at these higher frequencies. Multiple epochs can be used to improve the sensitivity. Ideally they should be aligned and combined before taking the log-amplitude. Combining aligned spectra reduces the noise before the nonlinear logarithm is taken, allowing Fourier structure that is noise-dominated in individual epochs to become measurable. By contrast, averaging the individual log-amplitudes improves the precision only over the frequency range accessible in each epoch. Still, when accurate alignment is unavailable, averaging the Fourier power spectra provides a shift-invariant alternative, provided that noise-dominated modes are excluded.

The ability to measure rotational broadening is determined by the leading terms that cannot be fully absorbed by the Voigt envelope. When rotational broadening is modest compared to the other velocity scales, the Fourier amplitude decays before reaching the first zero of the Bessel function. In this regime, the rotational contribution may be expanded in powers of $\zeta$,
\begin{equation}\label{eq: lnS approx}
\begin{aligned}
\ln|\tilde{S}| \simeq
& -\pi\,{v_\gamma}\,|\zeta| \\
& -\pi^2
\left[
\frac{v_{\textsc{g}}^2}{4\ln2}
-
\left(
\frac{\epsilon}{15}
-
\frac{1}{2}
\right)
v_{\textsc{r}}^2
\right]\zeta^2 \\
& - \pi^4
\left[
\left(
\frac{1}{24}
- \frac{\epsilon}{63}
\right)
v_{\textsc{r}}^4
\right]\zeta^4 
+
\mathrm{const},
\end{aligned}
\end{equation}
where the coefficients are quoted to leading order in the limb-darkening coefficient $\epsilon$. The constant, linear, and quadratic terms form the smooth Voigt-like envelope; the leading rotational information that survives this baseline is the $\zeta^4$ term. Appendix~\ref{app: expected precision} derives the corresponding Cram\'er--Rao bound, shown as black curves in the right panel of Figure~\ref{fig: line example}. 

This is the principle we reuse throughout: the measurable signal is the part of the log-amplitude that cannot be absorbed into the smooth envelope. A faint companion will be shown to enter at the same $\zeta^4$ order at small separation (see Section~\ref{sec: binary spec}), which is why the two are degenerate there.


\subsection{A faint companion}\label{sec: binary spec}

For binaries, the observed spectrum is the sum of the contributions from all surface elements of both stellar components, $\pspc{a}$ and $\pspc{b}$. Their respective kernels, $\pker{a}$ and $\pker{b}$, aggregate the flux from each component as measured in the observer's frame. The observed spectrum therefore consists of two terms,
\begin{equation}\label{eq: binary_system}
S = \pker{a} \circledast \pspc{a} + \pker{b} \circledast \pspc{b}.
\end{equation}
Here $\pspc{a}$ and $\pspc{b}$ denote the principal spectra of the primary and secondary stars, respectively, and $\pker{a}$ and $\pker{b}$ are their corresponding principal kernels. The equation above is equivalent to the \texttt{TODCOR} \citep{Zucker1994} modeling approach.

We now extend the formulation from the previous section to a superposition of two spectral lines. The Fourier-domain representation of the combined spectrum can then be written as
\begin{equation} \label{eq: lnS explicit}
\ln|\tilde{S}|
\simeq
\ln |\tpker{a}|
+
\ln\left|1+\alpha\,\mathcal{R}(\zeta)\,
e^{-i2\pi\zeta v}\right|
+\mathrm{const},
\end{equation}
where $v\equiv v_{\textsc{b}} - v_{\textsc{a}}$ is the instantaneous radial-velocity separation between the two stars. Here
\begin{equation*}
\alpha \equiv  \alpha_{\textsc{b}} / \alpha_{\textsc{a}}
\quad \text{and} \quad
\mathcal{R}(\zeta) \equiv {\tpker{b}}\big/{\tpker{a}},
\end{equation*}
where $\alpha_{\textsc{a}}$ and $\alpha_{\textsc{b}}$ are the corresponding line depths of the components and $\alpha$ denotes the ratio of the secondary to primary line amplitudes in the continuum-normalized spectrum. The ratio $\mathcal{R}$ measures how differently the two stars are broadened, and approaches unity for similar kernels.

For simplicity, we first consider the regime in which the instrumental line-spread function (LSF) dominates over the intrinsic line shapes.
Appendix~\ref{app: approx validity} outlines the validity of this assumption, and Section~\ref{sec: example phase fit} demonstrates the consequences of mild departures from it. In this regime, the two components have nearly the same effective kernel over the fitted Fourier range, so that $\mathcal{R}(\zeta)\simeq1$. This approximation is most appropriate when the intrinsic and rotational broadening of both components is small compared with the instrumental width, or when their effective line profiles are otherwise similar. Systems with strongly different rotational or atmospheric broadening, as may occur following binary interaction, require a more general treatment.

We further assume that the depth ratio is small and expand equation~(\ref{eq: lnS explicit}) to first order in $\alpha$. Both steps yield
\begin{equation}
\ln|\tilde{S}|
\approx
\ln|\tpker{a}|
+\alpha\,
\cos\left(2\pi \zeta v\right)
+\mathrm{const}.
\end{equation}
An appropriate model for this limiting case is
\begin{equation}\label{eq: unresolved binary model}
y_i = a_0 + a_1 \zeta_i+  a_2 \zeta_i^2
+ a_3 \cos(2\pi \zeta_i a_4)
+ {\rm noise},
\end{equation}
where $a_0$ describes the line-amplitude normalization, and $a_1$ and $a_2$ absorb the smooth Voigt component of the primary Fourier envelope. The coefficient $a_3$ estimates $\alpha$, while $a_4$ is the velocity separation and is the only non-linear parameter in this prescription. For multi-epoch data, $a_4$ is defined separately per epoch. 

When the primary velocity is known independently, the spectrum can be placed in the primary rest frame before taking the Fourier-domain logarithm. This retains phase information that is discarded in the amplitude-only expression. The linearized complex logarithm can be written as
\begin{equation}\label{eq: complex log approx}
   \ln(\tilde{S}_0) \approx \ln|\tilde{S}| - i\alpha \mathcal{R}(\zeta) \sin(2\pi\zeta v),
\end{equation}
where $\tilde{S}_0$ represents the shifted Fourier transformed signal and $\ln|\tilde{S}|$ was derived earlier. 
In an arbitrary reference frame, the phase also contains the primary's linear Doppler term. An independently known primary velocity allows this term to be removed; otherwise it must be fitted and can absorb the leading-order companion signal, as discussed below.
The complex phase can be modeled by taking the LSF-limited approximation as above,
\begin{equation}\label{eq: phase model}
    z_i = - b_0 \sin (2\pi\zeta_i b_1) + \text{noise},
\end{equation}
where $z_i \equiv \arg(\tilde{S}_{0,i})$, $b_0$ corresponds to $\alpha$, and $b_1$ represents the velocity shift. As in equation~\eqref{eq: single line model}, the additive noise term is a linearized approximation.

Figure~\ref{fig: double line example} illustrates the analysis of a synthetic composite line. The left panel shows the simulated spectrum, with the composite profile and the primary component indicated. The middle panel shows the Fourier amplitude of the composite line and its deviation from the single-lined baseline. The right panel shows the corresponding complex Fourier representation. The fitted Fourier amplitudes and phases are shown as thin solid red lines. The fit is restricted to conjugate wavelengths that are not dominated by noise, as indicated by the flattening of the log-spectrum in the middle panel. The dashed red lines show an extrapolation of the fit beyond this usable range, illustrating where the Fourier signal becomes noise dominated.

\section{Detection of faint companions}\label{sec: detection}
\subsection{Detection statistic}
The models derived above naturally define a statistical test for the presence of a faint companion. In the absence of a secondary component, the logarithmic Fourier amplitudes follow the smooth single-star model described in Section~\ref{sec: isolated lines}. A binary system introduces an additional modulation term whose structure depends on the regimes discussed in Section~\ref{sec: binary spec}.

The presence of a companion can therefore be assessed by comparing the likelihood of the data under the single-star model and under the binary model. Since both models are linear in the nuisance parameters describing the smooth baseline, these parameters can be optimized analytically for any fixed set of binary parameters.\footnote{This applies directly when fitting equation~\eqref{eq: unresolved binary model}, which neglects rotational broadening. Rotation can be included through the linear approximation in equation~\eqref{eq: lnS approx}, at the cost of a partial degeneracy with the companion signal at small velocity separations, since both enter at order $\zeta^4$ (Sections~\ref{sec: isolated lines} and~\ref{sec: detection efficiency}). When rotational broadening dominates,  $v_{\textsc{r}}$ must be fitted nonlinearly.} The resulting likelihood ratio, $\Lambda$, measures the improvement in the fit obtained when the additional modulation term is included \citep[cf.][]{Cash1979}. For Gaussian errors (see Appendix~\ref{app: noise model}) this ratio is equivalent to
\begin{equation*}
-2\ln\Lambda(v) =
\chi^2_{\textsc{h}_0} - \chi^2_{\textsc{h}_1}(v),
\end{equation*}
where $\chi^2_{\textsc{h}_0}$ is the minimum weighted residual sum obtained under the single-star model and $\chi^2_{\textsc{h}_1}(v)$ is the corresponding minimum obtained under the binary model for a given velocity separation. The detection statistic is obtained by maximizing this likelihood improvement over the allowed binary parameters,
\begin{equation} \label{eq: T statistic}
\mathcal{T} = \max_v \left[\chi^2_{\textsc{h}_0}-\chi^2_{\textsc{h}_1}(v)\right].
\end{equation}
Large values of $\mathcal{T}$ favor the binary interpretation. For absorption-line detection, negative fitted depth ratios are placed on the $\alpha=0$ boundary and assigned $\mathcal{T}=0$.

For regular nested models, Wilks' theorem gives an asymptotic chi-squared distribution for the likelihood-ratio statistic, with degrees of freedom equal to the number of additional fitted parameters \citep{Wilks1938}. The present search does not generally satisfy these conditions.
First, requiring $\alpha$ to be non-negative introduces a boundary, as discussed by \citet{Chernoff1954}.
Second, $\mathcal{T}$ is obtained by scanning over a parameter that is undefined under the null, since $v$ carries no meaning when the companion contribution vanishes. The maximization in equation~(\ref{eq: T statistic}) therefore places $\mathcal{T}$ in the regime discussed by \citet{Davies1977, Davies1987}.
We therefore calibrate its null distribution empirically.

\begin{figure}
    \includegraphics[width=0.97\linewidth]{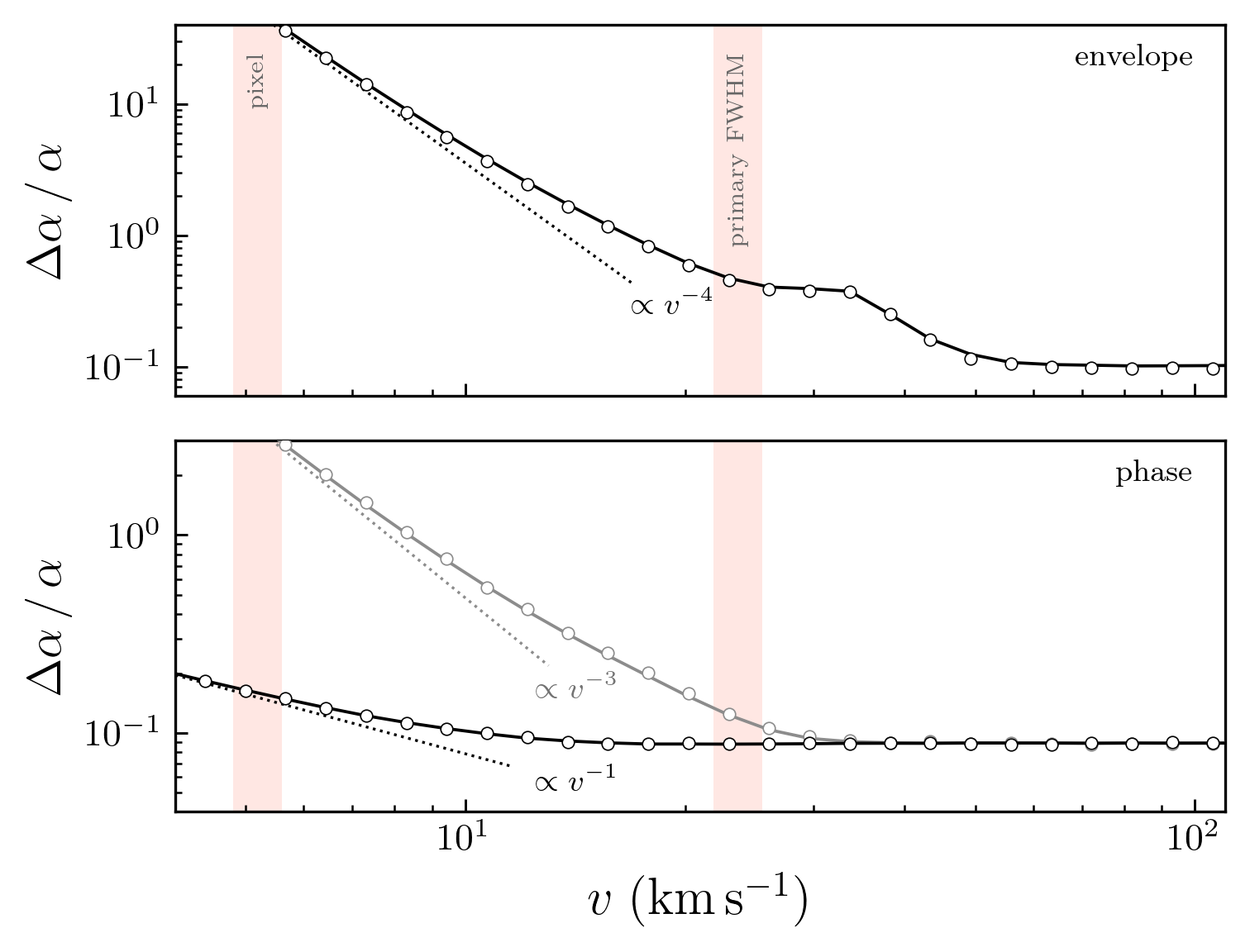}
\caption{Fractional depth-ratio error as a function of velocity separation for the two Fourier-domain channels. Both panels use the same parameters as the example in Figure~\ref{fig: double line example}; only the relative velocity $v$ is varied. At each plotted separation, $v$ is held fixed while $\alpha$ is fitted. The curves therefore show conditional precision and do not include the look-elsewhere effect associated with a velocity search. Top: log-amplitude channel. The analytic bound agrees with the Monte Carlo scatter and follows the expected $v^{-4}$ scaling at small separations. Bottom: phase channel. The black curves show the optimistic case in which the primary rest velocity is known, giving the $v^{-1}$ scaling. The gray curves show the result when the primary velocity is fitted. The fitted phase slope then absorbs the leading companion term, and the uncertainty follows the $v^{-3}$ scaling. Solid curves show the analytic bounds, open markers show the Monte Carlo scatter, and dotted lines mark the asymptotic scalings. The shaded vertical bands indicate the pixel sampling scale and the FWHM of the primary line. Values $\Delta\alpha/\alpha \gtrsim 1$ mark the practical non-detection.}
    \label{fig: precision curve}
\end{figure}

\begin{figure*}
    \centering
    \includegraphics[width=1\linewidth]{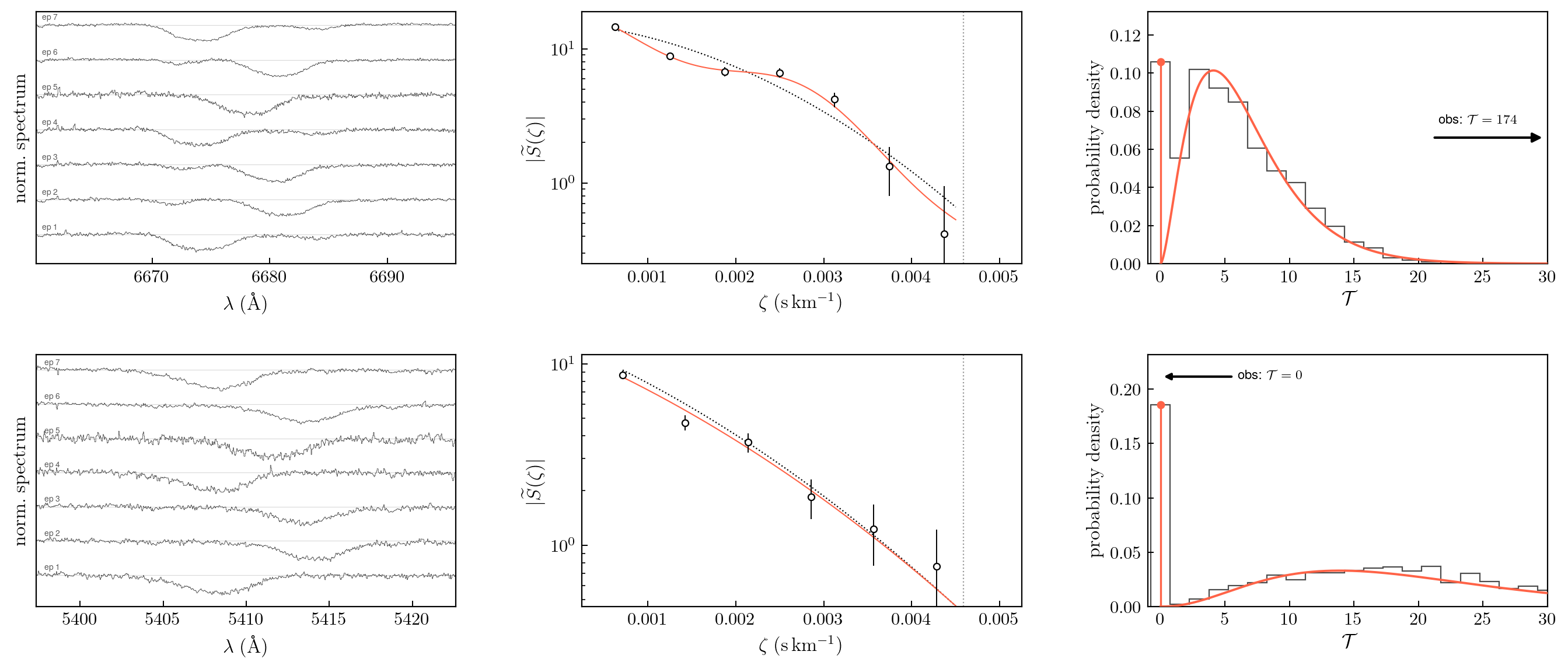}
    \caption{Companion-detection diagnostic applied to the FEROS spectra of CPD$-$41$^\circ$ 7742. The top row shows He\,I $6678~{ \AA}$, where both components are expected to contribute to the line profile. The bottom row shows He\,II $5411~{\rm \AA}$, which should be dominated by the primary. Left: locally normalized spectra from individual epochs, vertically offset. Middle: Fourier amplitudes used in the fit of the third epoch. The red curve is the best-fit two-component model, and the dotted curve is the corresponding single-star model obtained by setting $\alpha=0$ while keeping the fitted primary profile fixed. The vertical dotted line marks the Fourier-frequency cutoff used in the fit. The error bars include the empirical noise rescaling from the single-star injection--recovery test. Right: empirical null distribution of the detection statistic $\mathcal{T}$ (black histogram; 2500 single-star injection--recovery trials), used to compute an approximate tail probability under the single-star model. The fitted null distributions, defined in equation~(\ref{eq: chernoff-like}), are shown in red. The vertical stem marks the point-mass component at $\mathcal{T}=0$, corresponding to trials for which the best-fitting amplitude has $\alpha<0$. The continuous curve shows the Gamma distribution fitted to the positive-$\mathcal{T}$ tail, corresponding to trials with $\alpha>0$. The fitted values of the mixture weight and Gamma shape and scale are listed in Table~\ref{tab:feros_detection}. The arrow marks the observed value of $\mathcal{T}$. }
    \label{fig: FEROS example}
\end{figure*}

We use a \citeauthor{Chernoff1954}-like mixture model as a convenient description of the null distribution, with its parameters calibrated empirically. The calibration is based on injection--recovery tests in continuum regions adjacent to the analyzed line. In each trial, we inject a single synthetic line with shape parameters similar to those of the fitted primary line, place it at a random velocity and in a random continuum subsection, and analyze it with the same procedure used for the data. We fit both the single-star and binary models, scan over the velocity grid, and record the likelihood-ratio statistic $\mathcal{T}$. Trials for which the unconstrained best-fitting companion amplitude would be negative are placed on the $\alpha=0$ boundary and assigned $\mathcal{T}=0$. The resulting null distribution is approximated as
\begin{equation}\label{eq: chernoff-like}
\mathcal{T} \sim
w\,\delta_0 
+
(1-w)\,f\left(\mathcal{T}\right),
\end{equation}
where $w$ is the fraction of trials assigned to the $\mathcal{T}=0$ boundary, $\delta_0$ is a point mass at zero, and $f$ is the probability density of the positive-$\mathcal{T}$ trials. The latter is generally unknown and should be approximated or empirically quantified.

In the special case where all parameters are fixed and only the line-depth ratio is fitted, the null reduces to the \citet{Chernoff1954} result, with $w=0.5$ and $f$ approaching a $\chi^2$ distribution with a single degree of freedom. In the more general case, the positive-tail of the null, $f(\mathcal{T})$ in equation~\eqref{eq: chernoff-like}, is empirically derived from the positive injection--recovery trials. In the following we use a Gamma distribution to describe it. This is not a formal first-principles description of our expectations, but rather a flexible model that conveniently includes the $\chi^2$ family. It summarizes the combined effects of the velocity scan, correlations between neighboring trial velocities, and departures of the empirical noise from the ideal asymptotic assumptions.

A fitted null distribution is mainly useful for initial screening. Its parameters can be estimated from a relatively small Monte Carlo experiment, to identify clear cases for further inspection. When reliable constraints on the extreme tail of the null distribution are required, direct use of the injection--recovery distribution is preferable whenever sufficiently many trials are available. This avoids imposing a specific parametric form on the tail and naturally includes the look-elsewhere effect associated with the velocity scan. Alternatively, a more flexible tail model may be needed to adequately describe the null.

\subsection{Detection efficiency}\label{sec: detection efficiency}
The attainable precision on the line-depth ratio degrades steeply toward smaller velocity separations. In the LSF-limited regime, the fractional error on $\alpha$ inferred from the Fourier amplitudes of equation~(\ref{eq: unresolved binary model}) scales as
\begin{equation}\label{eq: alpha accuracy}
\frac{\Delta \alpha}{\alpha} \propto
\bigg(\frac{v_{\textsc{g}}}{v}\bigg)^4
\alpha_{\textsc{b}}^{-1}
\sigma\,,
\end{equation}
where $v_{\textsc{g}}$ represents the approximately Gaussian width of the line-spread function. This scaling reflects the loss of contrast between the two components as their velocities converge. In this limit, the companion-induced modulation is increasingly absorbed by the smooth single-star baseline, and the composite profile becomes difficult to distinguish from a single broadened line (see Appendix~\ref{app: expected precision}).

The top panel of Figure~\ref{fig: precision curve} illustrates this behavior for a simulated LSF-limited binary at a fixed known velocity separation. Although the perturbation is first order in the line-depth ratio $\alpha$, its leading measurable contribution after fitting the Voigt-like baseline is fourth order in the velocity separation. A similar degeneracy arises for marginally resolved rotational broadening. After the constant, linear, and quadratic terms of the Fourier-amplitude envelope are absorbed, the remaining rotational signature also enters at higher order. Thus, a severely blended double-lined binary observed at a single epoch can appear simply as a rotationally broadened line. The distinction is that rotational broadening is fixed in the stellar rest frame, whereas the blend-induced distortion changes with orbital phase. A static spectral blend within the analysis window, for example due to nebular or diffuse interstellar lines, is likewise a genuine departure from the isolated-single-line null and is therefore correctly flagged by the statistic. Multi-epoch behavior and astrophysical considerations are then required to distinguish such blends from companions.

\begin{figure*}
    \centering
    \includegraphics[width=1\linewidth]{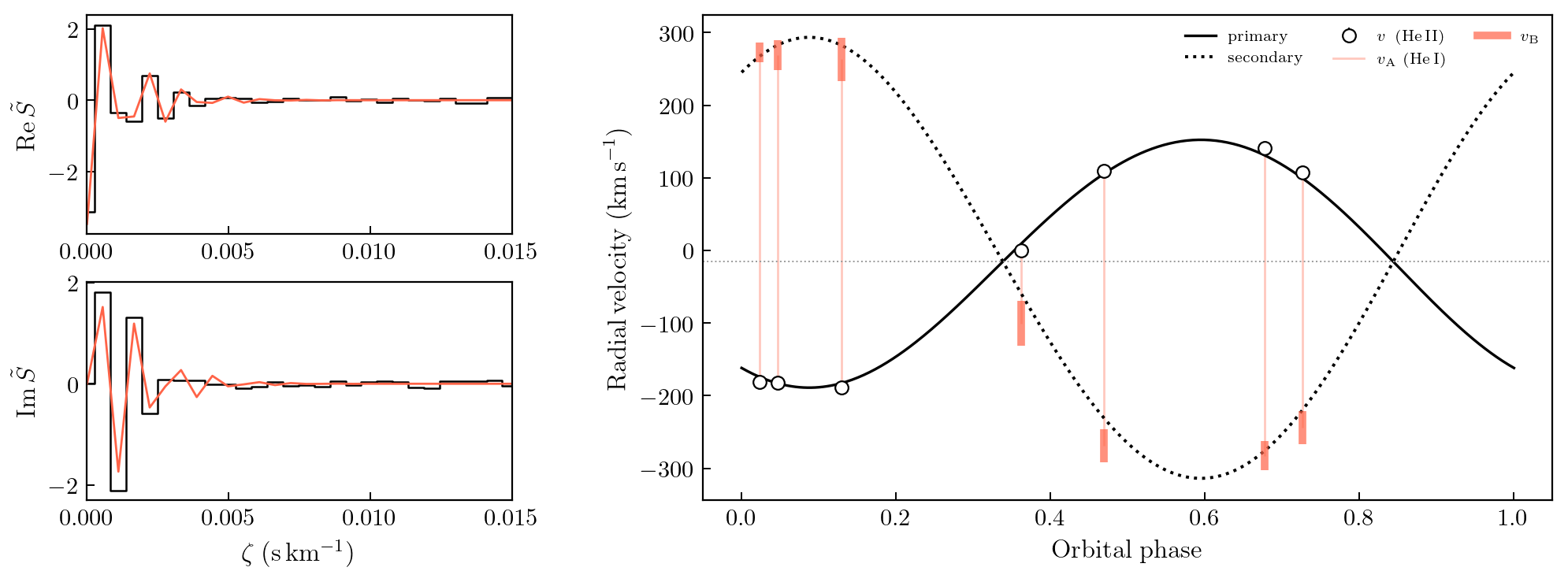}
    \caption{Epoch radial velocities of CPD$-$41$^\circ$ 7742. The left panels show the complex-phase fit to the isolated He~II line. Since this line is only produced by the primary star, the phase carries the Doppler shift signal. The right panel shows the primary velocities derived from the He~II line (open circles) with the velocity difference derived from the He~I line profile shape (red stripes). The amplitude fit determines the magnitude of the velocity separation, while the sign was assigned using the known orbital phase for this validation plot. The solid and dotted black lines represent the \citet{Rosu2022} orbital solution for reference.}
    \label{fig: binary orbit example}
\end{figure*}

A different scaling applies when detection is based on the Fourier phase, as described by equation~(\ref{eq: phase model}). In the linearized faint-companion limit, after shifting to the primary rest frame, the secondary contributes a sine-like phase term corresponding to an antisymmetric perturbation of the line profile about the primary line center. For $|\zeta v|\ll1$, this perturbation is linear in the velocity separation. Its precision follows from the same Fisher analysis as in Appendix~\ref{app: expected precision}, with the cosine template replaced by the sine term. The phase therefore responds to small line-profile asymmetries, whereas the Fourier amplitude is primarily sensitive to symmetric changes in the line shape.

Unlike the amplitude model, the phase model contains no smooth baseline that can absorb the leading-order companion signal. The signal therefore scales linearly with $v$, rather than at the higher order that remains after fitting the single-star amplitude envelope. When the orbital velocities are known independently, the fractional uncertainty scales as
\begin{equation}
\frac{\Delta\alpha}{\alpha}\propto v^{-1}.
\end{equation}
This optimistic limit is illustrated by the black curve in the lower panel of Figure~\ref{fig: precision curve}.

This scaling is conditional on a fixed trial velocity separation, as when an orbital solution is available. If $v$ is instead scanned, the leading phase signal depends on the product $\alpha v$. Although $\alpha$ and $v$ are not separately identified at this order, their product is sufficient for detecting a companion. The local sensitivity therefore retains the optimistic $v^{-1}$ scaling. The global detection threshold must, however, account for the look-elsewhere effect associated with the velocity scan. Separating $\alpha$ from $v$ requires the next-order sine term, for which the marginalized uncertainty on $\alpha$ scales as $v^{-3}$. The same $v^{-3}$ scaling arises when the primary rest velocity is fitted rather than known, since a linear-in-$\zeta$ nuisance term absorbs the leading phase signal in either case (gray curve in Figure~\ref{fig: precision curve}).

These asymptotic scalings apply in the perturbative, high-contrast regime. As stated above, interpreting the leading phase slope as a companion signature also requires an accurate primary velocity. An error in the adopted primary velocity produces the same phase slope as a faint companion, so the primary velocity must be constrained to better than the companion-induced shift, of order $\alpha v$. The approximation also breaks down for nearly identical twins. When the components have comparable line depths and similar intrinsic profiles, their unresolved composite changes mainly through symmetric broadening or narrowing, and assigning a unique primary velocity becomes partly degenerate with the secondary contribution. The phase-model sensitivity may then admit an extended locus of solutions in $(\alpha,v)$ rather than a single well-localized constraint.

In practice, the exact asymptotic behavior may differ from the idealized case because of imperfect line-profile modeling or continuum removal. Nevertheless, in the unresolved regime the sensitivity should still follow an approximate power-law trend,
\begin{equation}
\frac{\Delta \alpha}{\alpha} \propto v^{-\eta}.
\end{equation}
Here $\eta\,{\sim}\,4$ for the simple Gaussian LSF-limited Fourier-amplitude approach. For the phase channel, $\eta\,{\sim}\,1$ is the optimistic detection limit when the primary velocity is known, while $\eta\,{\sim}\,3$ applies when the primary velocity is fitted or when $\alpha$ and $v$ must be constrained separately.


\section{A worked example: CPD$-$41$^\circ$ 7742}\label{sec: worked example}
CPD$-$41$^\circ$ 7742 provides a non-trivial test case for the Fourier-domain companion search. It is a bright ($V \simeq 8.3$) eclipsing massive binary in NGC 6231 \citep{Baume1999}. The system contains two main-sequence stars, with spectral types O9 and B1, in a mildly eccentric ${\sim}\,2.4\,\mathrm{d}$ orbit \citep{Sana2003}. \citet{Rosu2022} recently presented a detailed spectroscopic analysis of both components. As discussed in Section~\ref{sec: example phase fit}, the primary is substantially more rotationally broadened than the secondary. The system therefore lies outside the regime in which the $\mathcal{R}\,{\simeq}\,1$ approximation is valid and provides a stringent test of the method's robustness to departures from it.

We use archival high-resolution spectra obtained with the Fiber-fed Extended Range Optical Spectrograph \citep[FEROS;][]{Kaufer_1998_FEROSNew, Kaufer_1999_CommissioningFEROS}, mounted on the MPG/ESO 2.2-meter telescope at La Silla Observatory. FEROS covers 350--920~nm at a resolving power of ${\sim}\,5\times10^4$, providing sufficient resolution and wavelength coverage to test the Fourier-domain binary signature in individual strong lines. Six epochs were obtained under ESO programme 0112.A-6007(A), with one additional epoch from 0111.A-9014(A). The typical per-pixel noise of the analyzed spectral regions is $\sigma_{\rm pix}{\sim}\,0.5-1.5\%$ in units of the normalized continuum.

We analyze two different spectral lines: He\,I $6678$~{\AA}, where the B-type secondary is expected to contribute, and He\,II $5411$~{\AA}, where the secondary contribution should be weak or absent  \citep[see the spectral disentangling by][]{Rosu2022}.  For each selected line, we fit the local continuum using wavelength intervals on both sides of the feature, divide the spectrum by this fit, and subtract unity so that the continuum level is zero.  We then apply a Tukey taper to the edges of the extracted window, reducing boundary discontinuities in the Fourier transform while leaving the central line profile unchanged \citep{Harris1978}. The noise is calibrated independently for each line by the injection--recovery procedure of Appendix~\ref{app: noise model}: we inject synthetic single-star profiles into nearby line-free regions of the same spectra, repeat the full preprocessing and fitting, and rescale the nominal Fourier-domain uncertainties so that the median $\chi^2$ of the single-star fits equals its number of degrees of freedom. This accounts for low-frequency structure in the FEROS spectra that is not captured by the pixel-level noise estimate. The error bars in the middle panels of Figure~\ref{fig: FEROS example} show these rescaled uncertainties.

We estimate the null distribution of the detection statistic, $\mathcal{T}$, using the same injection--recovery experiments. Each injection–recovery trial comprises seven synthetic single-star lines, one per epoch, injected at independently drawn per-epoch velocities, mimicking the primary's own epoch-to-epoch motion, into line-free continuum regions. Each trial is processed through the same joint seven-epoch fit and per-epoch velocity scan as applied to the data, so the null distribution includes the full look-elsewhere effect of the multi-epoch search. Each injected single-star line is fitted with both the single-star and binary models. The companion line-depth ratio is allowed to take either sign in the unconstrained fit; trials with a negative best-fitting ratio are subsequently assigned $\mathcal{T}=0$. The process is repeated 2500 times. We represent the resulting null distribution by a point mass at zero and a $\Gamma$-distribution fitted to the positive-$\mathcal{T}$ tail, according to equation~(\ref{eq: chernoff-like}). The Monte Carlo results and fitted null model are shown in the right panels of Figure~\ref{fig: FEROS example}.

\subsection{Blind Fourier-amplitude companion search}
We jointly fit all seven epochs using equation~(\ref{eq: unresolved binary model}), assuming $\mathcal{R}\simeq1$ and a Voigt-like primary line envelope. The primary-envelope coefficients and line-depth ratio ($a_0, a_1, a_2, a_3$) are fitted simultaneously, while the velocity separation ($a_4$) is allowed to vary independently between epochs and is searched over on a grid. To keep the fitted line profile physically sensible, the Lorentzian and Gaussian envelope coefficients ($a_1$ and $a_2$) are constrained to be non-positive. By contrast, the line-depth ratio ($a_3$) is left unconstrained in the linear fit. Although negative depth ratios of absorption lines are unphysical, they quantify the scatter of this parameter around zero. We therefore retain both the best-fitting value and its uncertainty, and use them to derive upper limits on the companion contribution when no significant positive signal is detected.

The top panels of Figure~\ref{fig: FEROS example} show the blind Fourier-amplitude search applied to He~I 6678~\AA. The left panel shows the analyzed spectral segment, where the secondary contribution is already visible in wavelength space. The middle panel shows the corresponding Fourier-amplitude fit: the binary model captures a clear oscillatory deviation from the smooth single-star envelope. The right panel shows the calibrated null distribution of the detection statistic. The stem at the origin represents injection--recovery trials in which the unconstrained companion amplitude was negative and was therefore placed on the $\alpha=0$ boundary. The continuous component describes the positive-$\mathcal{T}$ tail for non-negative $\alpha$. For He~I, the observed statistic lies beyond the plotted range, indicating an unambiguous detection of the secondary.

The bottom panels show the same blind Fourier-amplitude search applied to He~II 5411~\AA. In contrast to He~I 6678~\AA, the He~II line is dominated by the primary, with little or no contribution from the cooler secondary. The spectra show no obvious secondary feature in wavelength space, and the Fourier amplitudes remain consistent with a smooth single-star envelope. Across the velocity grid, the best-fitting line-depth ratios are negative, indicating that the cosine term is absorbing noise rather than a positive companion signal. Under the non-negativity boundary used for the detection test, these solutions are therefore placed at $\alpha=0$, and the corresponding detection statistic is set to $\mathcal{T}=0$. The He~II result thus lies on the point mass at the origin of the calibrated null distribution, yielding a non-detection rather than an independent companion detection.

\begin{table}
\centering
\caption{Companion-detection tests for CPD$-$41$^\circ$ 7742 using seven FEROS epochs. The top block shows the blind Fourier-amplitude search, in which the velocity separation is scanned independently at each epoch. The null distribution is calibrated empirically as a mixture of a point mass at zero and a Gamma-distributed positive component. The listed parameters are $(w,\mathrm{shape},\mathrm{scale})$, following the convention of \texttt{scipy.stats.gamma}. For He~I, the observed value of $\mathcal{T}$ exceeds those obtained in all 2500 injection--recovery trials. We therefore report the corresponding conservative empirical bound. The fitted Gamma distribution describes the bulk and moderate tail of the null distribution and is not extrapolated beyond the range sampled by the simulations. The bottom block shows the Fourier-phase search using the ephemeris-predicted velocities from \citet{Rosu2022}. In this case, the null distribution follows the \citet{Chernoff1954} mixture. The quantity $\hat{\alpha}$ denotes the fitted line-depth ratio. The quoted uncertainties are formal statistical errors computed using the empirically rescaled noise, without the interval-coverage correction of Appendix~\ref{app: noise model}. Upper limits are one-sided $95\%$ bounds from the positive tail of the Gaussian error distribution; for the amplitude search they are evaluated at the best-fitting velocities. For He~I, the phase-search significance is sufficiently high that the formal analytical \citeauthor{Chernoff1954} tail probability is numerically negligible.}
\label{tab:feros_detection}
\begin{tabular}{lcccc}
\hline\hline
Line & $\mathcal{T}$ & null parameters & $p$-value & $\hat{\alpha}$ (\%) \\[2pt]
\hline \noalign{\vskip 4pt}
He\,I  & 174 & $(0.15,2.7,2.4)$
       & $< 4\times10^{-4}$ & $24\pm2$ \\
He\,II & 0   &  $(0.25, 3.5,5.5)$
       & $\sim 1$ & $<2.1$ \\ \noalign{\vskip 4pt}
\hline \noalign{\vskip 4pt}
He\,I  & 169 & \citet{Chernoff1954}
       & $\ll 1\times 10^{-4}$ & $16\pm1$ \\
He\,II & 0   & \citet{Chernoff1954}
       & $\sim 1$ & $<1.1$ \\
\hline\hline
\end{tabular}
\end{table}

\subsection{Informed Fourier phase companion search} 

We next repeat the companion search using the Fourier phases, following equation~(\ref{eq: phase model}). In this fit, the velocities are not free parameters. We adopt the double-lined orbital solution of \citet{Rosu2022}: the primary velocities are used to construct $\tilde{S}_0$ by shifting each spectrum to the primary rest frame, while the predicted velocity separations fix the sine-wave template in each epoch. Assuming $\mathcal{R}\simeq1$, the only fitted parameter is therefore the line-depth ratio ($b_0$). The phase fit uses the same empirically calibrated noise scale as the amplitude search (Appendix~\ref{app: noise model}). Because the velocities are fixed and only $\alpha$ is fitted, there is no velocity scan, and the null distribution of the detection statistic is the \citet{Chernoff1954} mixture of Section~\ref{sec: detection}.

The phase fit differs from the blind Fourier-amplitude search in two respects. First, we adopt the literature orbital solution, fixing the primary and secondary velocities at each epoch rather than searching over them. This uses more information than the phase method strictly requires: accurate primary velocities alone would suffice for detection and for constraining the product $\alpha v$. In practice, however, adopting the full orbital solution eliminates the velocity-separation grid search discussed in Section~\ref{sec: detection efficiency}. The He~II-based primary velocities agree with the orbital solution, as shown in Section~\ref{sec: rv validation}, so the choice between these velocity estimates does not affect the result.

Second, as discussed above, there is no need to fit a Voigt-like envelope for the primary line. After shifting each epoch to the primary rest frame, we compare the residual Fourier phase directly with the sine modulation predicted for a faint companion. As in the amplitude analysis, the fitted linear coefficient $b_0$ is allowed to take negative values, so that its uncertainty reflects the scatter around zero. For the detection statistic, however, negative best-fitting amplitudes are placed on the $\alpha=0$ boundary and assigned $\mathcal{T}=0$. If the fitted amplitude is not significantly positive, we estimate a one-sided $95\%$ upper limit from the positive tail of the Gaussian error distribution.

\subsection{Line-depth ratio constraints}\label{sec: example phase fit}
The results of both schemes are listed in Table~\ref{tab:feros_detection}. The top two rows correspond to the blind Fourier-amplitude search, while the bottom two show the phase fit with fixed orbital velocities. For He~I~$6678$~\AA, the estimates differ at the ${\sim}\,3.5\sigma$ level, with ${\sim}\,24\%$ from the amplitude fit and ${\sim}\,16\%$ from the phase fit. The quoted uncertainties are formal statistical errors, computed with the rescaled noise but without the coverage correction of Appendix~\ref{app: noise model}, and do not include channel-dependent systematic effects. \citet{Rosu2022} reported an $18\pm3\%$ equivalent-width ratio. At the relevant mass ratio, solar-metallicity PoWR models \citep{Hainich2019} predict an equivalent-width ratio approximately $0.85$ times the rotation-free line-depth ratio, implying a line-depth ratio of $21\pm4\%$. This estimate lies between the two channels and is consistent with both. 

The difference between the Fourier estimates therefore indicates systematic effects that are not captured by their formal uncertainties. Besides the kernel mismatch discussed below, residual low-frequency structure from the local rectification contributes, to which the phase channel, which fits no baseline, is particularly sensitive. The same residual drives the He~II estimates negative in both channels. \citet{Rosu2022} measured projected rotational velocities of $\simeq146\,\mathrm{km\,s^{-1}}$ for the primary and $\simeq77\,\mathrm{km\,s^{-1}}$ for the secondary. The narrower secondary kernel decays more slowly in Fourier space, causing $\mathcal{R}$ to increase across the fitted range. This places the system in the less favorable configuration discussed in Appendix~\ref{app: approx validity}. The detection itself remains unambiguous, with $\mathcal{T}=174$ and $169$ for the amplitude and phase analyses, respectively. We therefore regard the amplitude fit as a robust detection diagnostic, while the spread between the two depth-ratio estimates provides a practical indication of the systematic uncertainty in $\alpha$, spanning $16$--$24\%$. This range is somewhat broader than that obtained from detailed spectral modeling, but follows from a simple and largely model-independent analysis. A more general implementation could fit the kernel ratio directly, which we defer to future work.

\subsection{Validation via radial-velocity extraction}\label{sec: rv validation}

As a consistency check, we show that the same Fourier representation can also be used to recover conventional radial velocities. We first measure the primary velocities from He~II 5411~\AA, where the system is effectively single-lined. We then combine these velocities with the velocity separations inferred from the He~I Fourier-amplitude fit to estimate the secondary velocities. The resulting velocities are compared with the orbital solution of \citet{Rosu2022}.

The He~II line profile is first determined by fitting equation~(\ref{eq: single line model}) to all exposures. The fitted profile parameters are then used as an estimate of the kernel in equation~(\ref{eq: sb1 Stilde}). With the line shape fixed, the only remaining free parameter in each epoch is the radial velocity, which enters through the Fourier phase:
\begin{equation}
  \tilde{S} \propto  \exp\left[-i2\pi\zeta(\lambda_0+v_\textsc{a})\right],
\end{equation}
as equation~(\ref{eq: sb1 Stilde}) shows. Since the line shape was determined, we fit a model of the form
\begin{equation}
y_i + i\,z_i=
-\hat{\alpha}_{\textsc{a}}\hat{U}(\zeta_i)
\exp\left[-i2\pi\zeta_i(c_0+c_1)\right] .
\end{equation}
Here $c_0$ absorbs the fixed line position, including the systemic velocity and any wavelength-zero-point offset, while $c_1$ is the instantaneous orbital shift of the primary.

An example of this phase fit is shown in the left panels of Figure~\ref{fig: binary orbit example}. The black step curves show the real and imaginary parts of the Fourier transform of He~II 5411~\AA, whose phase is modulated by the primary velocity. The thin red curves show the fitted model. The right panel shows the resulting primary velocities, together with secondary velocities inferred by adding the He~I velocity separations. Since equation~(\ref{eq: unresolved binary model}) determines the velocity separation only up to a sign, we assign the sign using the known orbital phase. The recovered velocities are consistent with the \citet{Rosu2022} orbital solution, demonstrating that the Fourier-domain companion search remains compatible with standard radial-velocity interpretation.

\section{From survey design to population inference}
\label{sec: selection function}
The detection scheme described above is linear in most of its fitted parameters. This makes repeated likelihood evaluations inexpensive and allows the detection probability to be calibrated directly as a function of line-depth ratio, velocity separation, and observing configuration. The resulting efficiency maps demonstrate how the sensitivity depends on spectral resolution, noise level, diagnostic line, and epoch coverage. We first illustrate this calculation for representative low- and high-resolution instruments. We then outline how atmosphere and orbital models translate the line-level sensitivity into constraints on physical companion models, and outline how the target-level likelihoods can be incorporated into population inference.

\subsection{Sensitivity in line-depth ratio and velocity separation}
The low-resolution case is representative of several large spectroscopic surveys, including SDSS-V \citep{Kollmeier2026}. At the resolving power of the BOSS spectrograph ($R\sim2000$; \citealt{Smee2013}), for example, the instrumental profile corresponds to a velocity scale of approximately $150~\mathrm{km\,s^{-1}}$, with a pixel spacing of approximately $70~\mathrm{km\,s^{-1}}$. The high-resolution case is more relevant to targeted follow-up of systems hosting compact-object candidates. As a representative example, we adopt UVES-like specifications ($R\sim40{,}000$; \citealt{Dekker2000}), with an oversampled Gaussian LSF having a characteristic velocity width of approximately $7.5~\mathrm{km\,s^{-1}}$ and a pixel spacing of approximately $1.5~\mathrm{km\,s^{-1}}$.

For a given line, noise level, velocity separation, and companion strength, we estimate the probability that the secondary would be detected. We simulate a composite absorption line with a continuum-normalized primary depth of $0.5$, rotational velocity broadening of $50~\mathrm{km\,s^{-1}}$, and Gaussian pixel noise of $\sigma_{\rm pix}=1\%$. At each assumed depth ratio and velocity separation, we generate synthetic binary spectra and apply the same preprocessing and Fourier-domain analysis used above. The corresponding single-star null distribution is calibrated through injection--recovery experiments. The resulting efficiency maps are single-epoch calculations for one diagnostic line and give the local detection probability at a specified instantaneous velocity separation. 

For a specified multi-epoch cadence, one spectrum is simulated for each planned epoch at the corresponding noise level and velocity separation. The joint log-likelihood is the sum of the epoch contributions, with a shared line-depth ratio and a separate velocity separation for each epoch. The separations may be fixed by an orbital model or searched independently, as in the seven-epoch analysis of Section~\ref{sec: worked example}. The null distribution and detection efficiency are calibrated by applying the same joint analysis to multi-epoch single-star and binary realizations. The gain therefore depends on the orbital phases and velocity separations sampled, rather than on the number of epochs alone. A systematic study of the dependence on the number of epochs and phase coverage is beyond the scope of this work.

\begin{figure}
    \includegraphics[width=0.9\linewidth]{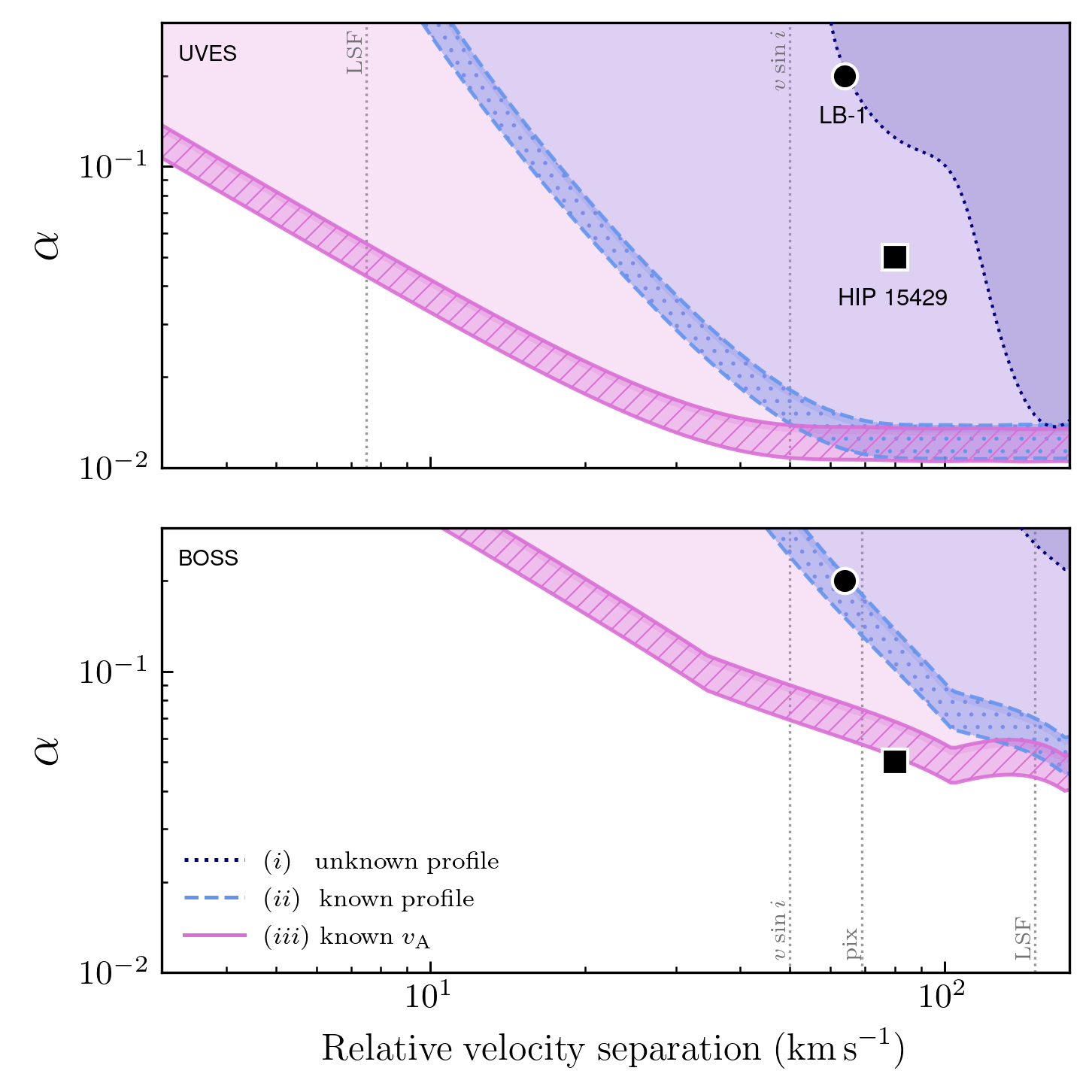}
\caption{Local companion-detection efficiency for UVES-like (top) and BOSS-like (bottom) observations of an isolated broadened line. The axes show the relative velocity separation and line-depth ratio. The curves show three idealized search configurations: (i) an amplitude-only search with an unknown primary profile, (ii) an amplitude search with a known primary profile, and (iii) a phase-sensitive search in which the primary radial velocity $v_{\rm A}$ is known. For each configuration, the critical value of the detection statistic is set by a $5\%$ false-alarm probability under the calibrated single-star null distribution. For configuration (i), the dotted line marks the $75\%$ detection-power contour. For configurations (ii) and (iii), the shaded bands indicate $50$--$75\%$ detection power, above which detection is highly probable. Vertical dotted lines mark the relevant instrumental and broadening velocity scales. LB-1 (\citealt{Abdul-Masih2020}) and HIP~15429 (\citealt{Muller-Horn2025}) are overplotted, using approximate quadrature velocity separations and representative He\,I line-depth ratios for the luminous secondary. These are merely illustrative benchmarks: the boundaries depend on the line depth and broadening.}
    \label{fig: power plot}
\end{figure}

We consider three idealized observing configurations. In case (i), no external information is assumed: the primary is represented by a Voigt profile, the line depth and depth ratio are fitted, and the velocity separation is scanned over. In case (ii), the full single-star line profile, including its rotational broadening, is assumed to be known; only the overall line depth, depth ratio, and velocity separation are fitted. In case (iii), the primary radial velocity is assumed to be known. Each spectrum is shifted to the primary rest frame before the Fourier phases are calculated, and the depth ratio is then fitted linearly at each trial velocity.

Figure~\ref{fig: power plot} compares the detection efficiency for the UVES-like and BOSS-like configurations. The dotted dark-blue contour and the dashed blue and solid magenta bands correspond, respectively, to an unconstrained amplitude search, a search with a known primary profile, and a phase-based search with a known primary radial velocity. For each case, the threshold is set at a $5\%$ false-alarm probability. The bands span detection powers of $50$--$75\%$, and the dotted contour marks $75\%$ power for the unconstrained search, whose transition is too steep to display as a band. They therefore mark gradual sensitivity boundaries rather than sharp limits. The resulting sensitivity map is expressed in the directly measured parameter space, $(\alpha,v)$, and provides the natural starting point for survey planning. Given an instrumental setup, diagnostic line, noise level, and observing cadence, it identifies the combinations of line-depth ratio and instantaneous velocity separation for which a companion is likely to be detected. Equally importantly, it identifies the domain in which a non-detection carries astrophysical meaning. The efficiency maps in Figure~\ref{fig: power plot} are computed under idealized noise conditions. Their empirical validation on real spectra, via companion injection into survey data, is an essential step in realistic applications and is deferred to dedicated work.

\begin{figure}
    \includegraphics[width=0.9\linewidth]{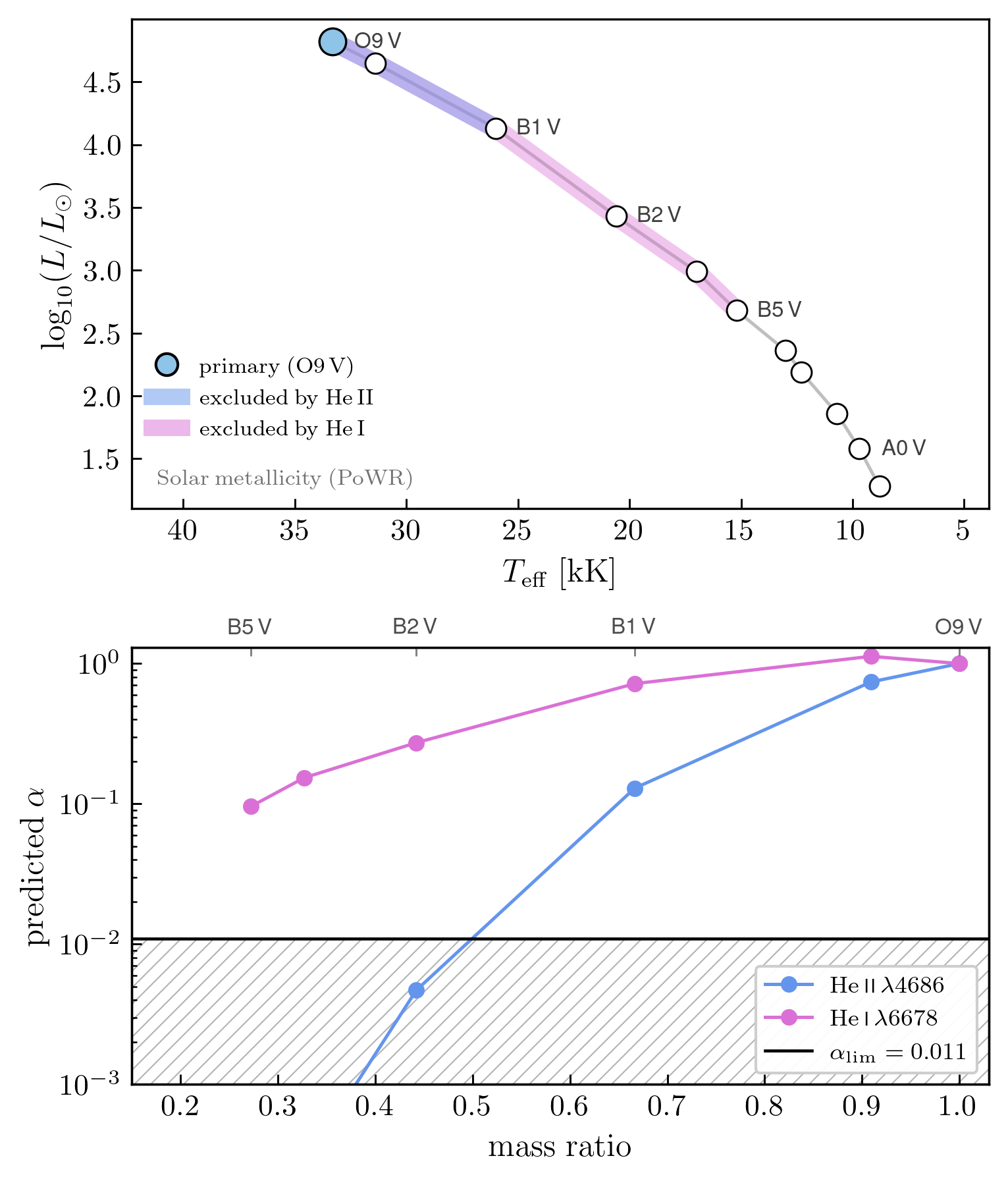}
\caption{Illustrative translation of the phase-channel upper limit $\alpha < 1.1\%$ into constraints on coeval, non-interacting main-sequence companions to an O9\,V primary. Top: main-sequence locus and the portions for which the predicted He~I~$6678$~\AA~or He~II~$4686$~\AA~line contribution exceeds the adopted detection limit. Bottom: predicted secondary-to-primary line-depth ratios, including continuum dilution, as a function of mass ratio. The horizontal line marks the observational upper limit; companions above it would have produced a detectable contribution under the assumed line models and observing configuration.}
    \label{fig: cmd exlusion}
\end{figure}

More generally, converting the observational sensitivity into a population-level selection function requires connecting $(\alpha ,v)$ to the physical binary parameters. 
Figure~\ref{fig: cmd exlusion} illustrates the next step in this interpretation. We use PoWR atmosphere models \citep{Hainich2019} to translate an upper limit of $\alpha<1.1\%$ into constraints on the properties of a main-sequence companion to a primary similar to CPD$-$41$^\circ$7742. Because different ionization stages respond differently to the secondary temperature and luminosity, the corresponding limits are evaluated using both diagnostic ions. For illustration, the same limit is applied to one representative line of each ion, He~I~6678~\AA\ and He~II~4686~\AA. The figure therefore shows how a line-level non-detection can exclude portions of the companion parameter space, rather than merely indicating that no second line was visually identified. The calculation shown here is restricted to non-interacting main-sequence binaries, but the same procedure can be extended to other evolutionary states.

\subsection{Prospects for population inference}

The upper limits illustrated in Figure~\ref{fig: cmd exlusion} provide a simplified example of constraints on an individual system. More generally, stellar-atmosphere models map the physical properties of a possible companion onto the expected line-depth ratios, while the orbital solution determines the velocity separations at the observed epochs. Together, these quantities specify the predicted line profiles. The likelihood developed here quantifies the probability of the observed spectra for each such companion model.

In this formulation, a non-detection is informative. Companion models that predict secondary lines inconsistent with the observed spectra have low likelihood and are therefore disfavored. In searches for quiescent black holes, these spectroscopic constraints complement the dynamical mass function by excluding luminous companions allowed by the orbital solution. A full Bayesian analysis would combine the spectroscopic likelihood with prior information about the orbit, stellar atmospheres, and the possible companion classes to obtain their posterior probabilities. The development and validation of this framework are deferred to future work.

The discussion so far has concerned individual systems. At the survey level, the target-level likelihoods can be combined within a hierarchical population model that accounts for variations in stellar properties, data quality, diagnostic lines, and epoch coverage. This provides the spectroscopic component required to interpret a sample of dark-companion candidates and infer the incidence of black-hole companions to hot stars.

\section{Summary}
\label{sec:summary}

Searches for dark companions to hot stars require quantitative constraints on faint luminous companions. A system may appear single-lined even when the companion is luminous enough to mimic a compact object dynamically. A spectral non-detection is therefore meaningful only when accompanied by a quantitative estimate of the luminous companions to which the observations were sensitive.

We used a simple line-profile approximation to derive such limits from isolated, broadened spectral lines. In Fourier space, a single-star line is represented by a smooth amplitude envelope, while a companion introduces additional structure that depends on its velocity separation and line shape. This defines a likelihood-ratio test between a smooth single-star model and a binary model, with the nuisance parameters of the primary line profiled out. The discussion above is focused on absorption lines; therefore, $\alpha$ is strictly non-negative. However, emission lines can yield negative depth ratios. The analysis also assumes that phase variations arise only from Doppler shifts. Line-profile deformations caused by pulsations, surface-temperature variations, winds, or instrumental effects may also alter the phase; their treatment within the broader formalism of \citet{Shahaf2023,Shahaf2026} is deferred to future work.

The analysis shows that companion detectability is controlled by both signal-to-noise and line-profile degeneracy. At small velocity separations, the companion signature can be absorbed by the single-star model, producing a power-law loss of sensitivity. Fourier amplitudes provide a direct test of line-profile structure. Fourier phases can add sensitivity when the primary velocities, or the wavelength zero point, are known independently and accurately enough. We validated the approach with archival FEROS spectra of CPD$-41^\circ$~7742. The method detects the known secondary in He\,I $6678\, {\rm \AA}$, where both stars contribute, and gives a stringent non-detection in He\,II $5411 \, {\rm \AA}$, where the cooler secondary should be weak or absent. This line-by-line behavior is the desired outcome for a practical companion diagnostic.
The same calculation gives local detection efficiencies as a function of line-depth ratio, velocity separation, noise level, and observing configuration. These efficiencies can be combined over lines, epochs, and targets to build survey selection functions. The present implementation is limited to isolated lines and approximate line-profile models, and its simplified cosine form is most appropriate when the two effective kernels are similar. Applications to Gaia RVS will require a dedicated treatment of the Paschen-series line structure, sampling, line-spread function, and epoch distribution. 

It is important to emphasize that the Fourier-domain analysis does not create information. The Fourier transform itself preserves the information in the spectrum, but taking the amplitude and truncating the Fourier range discards some of it. The method trades this information for conceptual and computational simplicity. Compared with full spectral disentangling, which reconstructs the component spectra across the wavelength grid, this approach fits only a small set of line-profile and companion parameters. Reducing the number of nuisance parameters can improve the separation between the null and alternative hypotheses. The resulting linear models are also fast to solve and are less susceptible to local minima and pathological solutions.

The method is motivated in particular by searches for black-hole companions to hot stars. Its purpose is nevertheless more general: to turn spectra into calibrated physical constraints that support both object-level classification and population-level inference. It is not a replacement for detailed spectral modeling. A sufficiently accurate physical model of the line shape, including mild asymmetries, may provide greater discriminating power. Its main advantage lies instead in broad and controlled searches. The formalism gives transparent estimates of the precision with which rotation or pressure broadening can be measured, allowing detection boundaries to be calculated before a survey is conducted. It also provides a simple and reproducible search procedure with few user-defined choices. In some cases, its upper limits may be weaker than those from a detailed model of an individual system, but their uniformity makes them useful for population-level inference.

 



\begin{acknowledgements}
We thank the referee for the thorough and constructive review of the manuscript. S. Shahaf, H. W. Rix, J. M\"uller-Horn, and J. I. Villase\~nor acknowledge support from the European Research Council for the ERC Advanced Grant [101054731]. This research was supported in part by grant NSF PHY-2309135 to the Kavli Institute for Theoretical Physics (KITP). TS acknowledges support from the European Research Council (ERC) under the European Union's Horizon 2020 research and innovation program (grant agreement 101164755/METAL) and the Israel Science Foundation (ISF) under grant number 0603225041.
\end{acknowledgements}

%

%
\bibliographystyle{aa}
\bibliography{hsb.bib}

\begin{thebibliography}{51}
\expandafter\ifx\csname natexlab\endcsname\relax\def\natexlab#1{#1}\fi

\bibitem[{{Abdul-Masih} {et~al.}(2020){Abdul-Masih}, {Banyard}, {Bodensteiner},
  {Bowman}, {Dsilva}, {Fabry}, {Hawcroft}, {Mahy}, {Marchant}, {Raskin},
  {Reggiani}, {Shenar}, {Tkachenko}, {Van Winckel}, \&
  {Sana}}]{Abdul-Masih2020}
{Abdul-Masih}, M., {Banyard}, G., {Bodensteiner}, J., {et~al.} 2020, \nat, 580,
  E11

\bibitem[{{Angelo} {et~al.}(2026){Angelo}, {Petigura}, \&
  {Bedell}}]{Angelo2026}
{Angelo}, I., {Petigura}, E., \& {Bedell}, M. 2026, \apj, 997, 238

\bibitem[{{Baume} {et~al.}(1999){Baume}, {V{\'a}zquez}, \&
  {Feinstein}}]{Baume1999}
{Baume}, G., {V{\'a}zquez}, R.~A., \& {Feinstein}, A. 1999, \aaps, 137, 233

\bibitem[{{Binnenfeld} {et~al.}(2025){Binnenfeld}, {Lilek}, {Nasser}, {Giryes},
  \& {Zucker}}]{Binnenfeld2025}
{Binnenfeld}, A., {Lilek}, S., {Nasser}, R., {Giryes}, R., \& {Zucker}, S.
  2025, \aap, 699, A294

\bibitem[{{Bodensteiner} {et~al.}(2022){Bodensteiner}, {Heida}, {Abdul-Masih},
  {Baade}, {Banyard}, {Bowman}, {Fabry}, {Frost}, {Mahy}, {Marchant},
  {M{\'e}rand}, {Reggiani}, {Rivinius}, {Sana}, {Selman}, \&
  {Shenar}}]{Bodensteiner2022}
{Bodensteiner}, J., {Heida}, M., {Abdul-Masih}, M., {et~al.} 2022, The
  Messenger, 186, 3

\bibitem[{{Bodensteiner} {et~al.}(2020){Bodensteiner}, {Shenar}, {Mahy},
  {Fabry}, {Marchant}, {Abdul-Masih}, {Banyard}, {Bowman}, {Dsilva}, {Frost},
  {Hawcroft}, {Reggiani}, \& {Sana}}]{Bodensteiner2020}
{Bodensteiner}, J., {Shenar}, T., {Mahy}, L., {et~al.} 2020, \aap, 641, A43

\bibitem[{{Cash}(1979)}]{Cash1979}
{Cash}, W. 1979, \apj, 228, 939

\bibitem[{Chernoff(1954)}]{Chernoff1954}
Chernoff, H. 1954, The Annals of Mathematical Statistics, 573

\bibitem[{Davies(1977)}]{Davies1977}
Davies, R.~B. 1977, Biometrika, 64, 247

\bibitem[{Davies(1987)}]{Davies1987}
Davies, R.~B. 1987, Biometrika, 74, 33

\bibitem[{{Dekker} {et~al.}(2000){Dekker}, {D'Odorico}, {Kaufer}, {Delabre}, \&
  {Kotzlowski}}]{Dekker2000}
{Dekker}, H., {D'Odorico}, S., {Kaufer}, A., {Delabre}, B., \& {Kotzlowski}, H.
  2000, in Society of Photo-Optical Instrumentation Engineers (SPIE) Conference
  Series, Vol. 4008, Optical and IR Telescope Instrumentation and Detectors,
  ed. M.~{Iye} \& A.~F. {Moorwood}, 534--545

\bibitem[{{El-Badry} \& {Quataert}(2021)}]{El-Badry2021}
{El-Badry}, K. \& {Quataert}, E. 2021, \mnras, 502, 3436

\bibitem[{{Essick} \& {Farah}(2026)}]{Essick2026}
{Essick}, R. \& {Farah}, A.~M. 2026, arXiv e-prints, arXiv:2605.06636

\bibitem[{{Fryer} {et~al.}(2012){Fryer}, {Belczynski}, {Wiktorowicz},
  {Dominik}, {Kalogera}, \& {Holz}}]{Fryer2012}
{Fryer}, C.~L., {Belczynski}, K., {Wiktorowicz}, G., {et~al.} 2012, \apj, 749,
  91

\bibitem[{{Giacobbo} \& {Mapelli}(2018)}]{Giacobbo2018}
{Giacobbo}, N. \& {Mapelli}, M. 2018, \mnras, 480, 2011

\bibitem[{{Gonz{\'a}lez} \& {Levato}(2006)}]{Gonzalez2006}
{Gonz{\'a}lez}, J.~F. \& {Levato}, H. 2006, \aap, 448, 283

\bibitem[{{Gonz{\'a}lez} {et~al.}(2024){Gonz{\'a}lez}, {Mart{\'\i}nez}, \&
  {Alejo}}]{Gonzalez2024}
{Gonz{\'a}lez}, J.~F., {Mart{\'\i}nez}, C.~I., \& {Alejo}, A. 2024, \aap, 690,
  A124

\bibitem[{{Gray}(1973)}]{Gray1973}
{Gray}, D.~F. 1973, \apj, 184, 461

\bibitem[{{Gray}(1992)}]{Gray1992}
{Gray}, D.~F. 1992, {The observation and analysis of stellar photospheres.},
  Vol.~20

\bibitem[{{Hainich} {et~al.}(2019){Hainich}, {Ramachandran}, {Shenar},
  {Sander}, {Todt}, {Gruner}, {Oskinova}, \& {Hamann}}]{Hainich2019}
{Hainich}, R., {Ramachandran}, V., {Shenar}, T., {et~al.} 2019, \aap, 621, A85

\bibitem[{Harris(1978)}]{Harris1978}
Harris, F. 1978, Proceedings of the IEEE, 66, 51

\bibitem[{Kaufer \& Pasquini(1998)}]{Kaufer_1998_FEROSNew}
Kaufer, A. \& Pasquini, L. 1998, in Optical {{Astronomical Instrumentation}},
  Vol. 3355 (SPIE), 844--854

\bibitem[{Kaufer {et~al.}(1999)Kaufer, Stahl, Tubbesing, N{\o}rregaard, Avila,
  Francois, Pasquini, \& Pizzella}]{Kaufer_1999_CommissioningFEROS}
Kaufer, A., Stahl, O., Tubbesing, S., {et~al.} 1999, The Messenger, 95, 8

\bibitem[{{Kollmeier} {et~al.}(2026){Kollmeier}, {Rix}, {Aerts}, {Aird}, {Vera
  Alfaro}, {Almeida}, {Anderson}, {Arseneau}, {Assef}, {Aviram}, {Aydar},
  {Badenes}, {Bandyopadhyay}, {Barger}, {Barkhouser}, {Bauer}, {Behmard},
  {Bender}, {Besser}, {Bhattarai}, {Bilgi}, {Bird}, {Bizyaev}, {Blanc},
  {Blanton}, {Bochanski}, {Bovy}, {Brandon}, {Brandt}, {Brownstein}, {Buchner},
  {Burchett}, {Carlberg}, {Casey}, {Castaneda-Carlos}, {Chakraborty},
  {Chanam{\'e}}, {Chandra}, {Cherinka}, {Chilingarian}, {Comparat}, {Cosens},
  {Covey}, {Crane}, {Crumpler}, {Cruz-Gonzalez}, {Cunha}, {Cunningham}, {Dai},
  {Darling}, {Davidson}, {Davis}, {De Lee}, {Deacon}, {M{\'e}ndez Delgado},
  {Demasi}, {Demianenko}, {Derwent}, {D'Onghia}, {Di Mille}, {Dias}, {Donor},
  {Dow}, {Drory}, {Dwelly}, {Egorov}, {Egorova}, {El-Badry}, {Engelman},
  {Eracleous}, {Fan}, {Farr}, {Fries}, {Frinchaboy}, {Froning}, {G{\"a}nsicke},
  {Garc{\'\i}a}, {Gelfand}, {Gentile Fusillo}, {Glover}, {Grabowski}, {Grebel},
  {Green}, {Grier}, {Gupta}, {Gray}, {H{\"a}berle}, {Hall}, {Hammond},
  {Hawkins}, {Harding}, {Heged{\H{u}}s}, {Herbst}, {Hermes}, {Rodr{\'\i}guez
  Hidalgo}, {Hilder}, {Hogg}, {Holtzman}, {Horta}, {Huang}, {Hwang},
  {Ibarra-Medel}, {Imig}, {Inight}, {Jana}, {Ji}, {Jim{\'e}nez-Arranz},
  {Jofre}, {Johns}, {Johnson}, {Johnson}, {Johnston}, {Jones}, {Katkov},
  {Knapp}, {Koekemoer}, {Kounkel}, {Kreckel}, {Krishnarao}, {Krumpe}, {Kumari},
  {Kupfer}, {Lacerna}, {Laporte}, {Lepine}, {Li}, {Liu}, {Loebman}, {Long},
  {Roman-Lopes}, {Lu}, {Majewski}, {Maoz}, {McKinnon}, {Medan}, {Merloni},
  {Minniti}, {Morrison}, {Myers}, {M{\'e}sz{\'a}ros}, {Nandra}, {Nayak},
  {Ness}, {Nidever}, {O'Brien}, {Oeur}, {Oravetz}, {Oravetz}, {Otto},
  {Pallathadka}, {Palunas}, {Pan}, {Pappalardo}, {Pandey}, {Negrete
  Pe{\~n}aloza}, {Pinsonneault}, {Pogge}, {Taghizadeh Popp}, {Price-Whelan},
  {Pulatova}, {Qiu}, {Ramirez}, {Rankine}, {Ricci}, {Runnoe}, {Sanchez},
  {Salvato}, {Sarbadhicary}, {Sattler}, {Saydjari}, {Sayres}, {Schinnerer},
  {Schlaufman}, {Schneider}, {Schreiber}, {Schwope}, {Serna}, {Shen},
  {Sif{\'o}n}, {Singh}, {Sinha}, {Smee}, {Song}, {Souto}, {Stassun},
  {Steinmetz}, {Stone-Martinez}, {Stringfellow}, {Stutz},
  {S{\'a}nchez-Gallego}, {Tan}, {Tayar}, {Thai}, {Thakar}, {Ting}, {Tkachenko},
  {Tovmassian}, {Trakhtenbrot}, {Fern{\'a}ndez-Trincado}, \&
  {Troup}}]{Kollmeier2026}
{Kollmeier}, J.~A., {Rix}, H.-W., {Aerts}, C., {et~al.} 2026, \aj, 171, 52

\bibitem[{{Langer} {et~al.}(2020){Langer}, {Sch{\"u}rmann}, {Stoll},
  {Marchant}, {Lennon}, {Mahy}, {de Mink}, {Quast}, {Riedel}, {Sana},
  {Schneider}, {Schootemeijer}, {Wang}, {Almeida}, {Bestenlehner},
  {Bodensteiner}, {Castro}, {Clark}, {Crowther}, {Dufton}, {Evans}, {Fossati},
  {Gr{\"a}fener}, {Grassitelli}, {Grin}, {Hastings}, {Herrero}, {de Koter},
  {Menon}, {Patrick}, {Puls}, {Renzo}, {Sander}, {Schneider}, {Sen}, {Shenar},
  {Sim{\'o}n-D{\'\i}as}, {Tauris}, {Tramper}, {Vink}, \& {Xu}}]{Langer2020}
{Langer}, N., {Sch{\"u}rmann}, C., {Stoll}, K., {et~al.} 2020, \aap, 638, A39

\bibitem[{{Liu} {et~al.}(2019){Liu}, {Zhang}, {Howard}, {Bai}, {Lu}, {Soria},
  {Justham}, {Li}, {Zheng}, {Wang}, {et~al.}}]{Liu2019}
{Liu}, J., {Zhang}, H., {Howard}, A.~W., {et~al.} 2019, \nat, 575, 618

\bibitem[{{Mahy} {et~al.}(2022){Mahy}, {Sana}, {Shenar}, {Sen}, {Langer},
  {Marchant}, {Abdul-Masih}, {Banyard}, {Bodensteiner}, {Bowman}, {Dsilva},
  {Fabry}, {Hawcroft}, {Janssens}, {Van Reeth}, \& {Eldridge}}]{Mahy2022}
{Mahy}, L., {Sana}, H., {Shenar}, T., {et~al.} 2022, \aap, 664, A159

\bibitem[{{Marchant} \& {Bodensteiner}(2024)}]{Marchant2024}
{Marchant}, P. \& {Bodensteiner}, J. 2024, \araa, 62, 21

\bibitem[{{M{\"u}ller-Horn} {et~al.}(2025){M{\"u}ller-Horn}, {El-Badry}, {Rix},
  {Shenar}, {Seeburger}, {Villase{\~n}or}, {Bodensteiner}, {Latham}, {Bieryla},
  {Buchhave}, {Isaacson}, \& {Howard}}]{Muller-Horn2025}
{M{\"u}ller-Horn}, J., {El-Badry}, K., {Rix}, H.-W., {et~al.} 2025, \aap, 701,
  A9

\bibitem[{{M{\"u}ller-Horn} {et~al.}(2026){M{\"u}ller-Horn}, {Ramachandran},
  {El-Badry}, {Sander}, {Bodensteiner}, {Gies}, {G{\"o}tberg}, {Rivinius},
  {Shenar}, {Sch{\"o}sser}, {Wang}, {Bieryla}, {Buchhave}, \&
  {Latham}}]{Muller-Horn2026}
{M{\"u}ller-Horn}, J., {Ramachandran}, V., {El-Badry}, K., {et~al.} 2026, \aap,
  708, A187

\bibitem[{{Ness} {et~al.}(2015){Ness}, {Hogg}, {Rix}, {Ho}, \&
  {Zasowski}}]{Ness2015}
{Ness}, M., {Hogg}, D.~W., {Rix}, H.-W., {Ho}, A. Y.~Q., \& {Zasowski}, G.
  2015, \apj, 808, 16

\bibitem[{{Rix} {et~al.}(2021){Rix}, {Hogg}, {Boubert}, {Brown}, {Casey},
  {Drimmel}, {Everall}, {Fouesneau}, \& {Price-Whelan}}]{Rix2021}
{Rix}, H.-W., {Hogg}, D.~W., {Boubert}, D., {et~al.} 2021, \aj, 162, 142

\bibitem[{{Rosu} {et~al.}(2022){Rosu}, {Rauw}, {Naz{\'e}}, {Gosset}, \&
  {Sterken}}]{Rosu2022}
{Rosu}, S., {Rauw}, G., {Naz{\'e}}, Y., {Gosset}, E., \& {Sterken}, C. 2022,
  \aap, 664, A98

\bibitem[{{Rowan} {et~al.}(2026){Rowan}, {Kraus}, \& {Thompson}}]{Rowan2026}
{Rowan}, D.~M., {Kraus}, S., \& {Thompson}, T.~A. 2026, \pasp, 138, 024203

\bibitem[{{Rowan} {et~al.}(2025){Rowan}, {Thompson}, {Kochanek}, {Li Causi},
  {Roth}, {Vaccari}, {Pedichini}, {Piazzesi}, {Antoniucci}, {Testa}, {Johnson},
  {Crass}, {Crepp}, {Bechter}, {Bechter}, {Sands}, {Harris}, \&
  {Fulton}}]{Rowan2025}
{Rowan}, D.~M., {Thompson}, T.~A., {Kochanek}, C.~S., {et~al.} 2025, \apj, 981,
  94

\bibitem[{{Royer} {et~al.}(2002){Royer}, {Gerbaldi}, {Faraggiana}, \&
  {G{\'o}mez}}]{Royer2002}
{Royer}, F., {Gerbaldi}, M., {Faraggiana}, R., \& {G{\'o}mez}, A.~E. 2002,
  \aap, 381, 105

\bibitem[{{Sana} {et~al.}(2003){Sana}, {Hensberge}, {Rauw}, \&
  {Gosset}}]{Sana2003}
{Sana}, H., {Hensberge}, H., {Rauw}, G., \& {Gosset}, E. 2003, \aap, 405, 1063

\bibitem[{{Seeburger} {et~al.}(2024){Seeburger}, {Rix}, {El-Badry}, {Xiang}, \&
  {Fouesneau}}]{Seeburger2024}
{Seeburger}, R., {Rix}, H.-W., {El-Badry}, K., {Xiang}, M., \& {Fouesneau}, M.
  2024, \mnras, 530, 1935

\bibitem[{{Shahaf} \& {Zackay}(2023)}]{Shahaf2023}
{Shahaf}, S. \& {Zackay}, B. 2023, \mnras, 525, 6223

\bibitem[{{Shahaf} \& {Zackay}(2026)}]{Shahaf2026}
{Shahaf}, S. \& {Zackay}, B. 2026, \mnras, 547, stag334

\bibitem[{{Shenar} {et~al.}(2020){Shenar}, {Bodensteiner}, {Abdul-Masih},
  {Fabry}, {Mahy}, {Marchant}, {Banyard}, {Bowman}, {Dsilva}, {Hawcroft},
  {Reggiani}, \& {Sana}}]{Shenar2020}
{Shenar}, T., {Bodensteiner}, J., {Abdul-Masih}, M., {et~al.} 2020, \aap, 639,
  L6

\bibitem[{{Shenar} {et~al.}(2022){Shenar}, {Sana}, {Mahy}, {El-Badry},
  {Marchant}, {Langer}, {Hawcroft}, {Fabry}, {Sen}, {Almeida}, {Abdul-Masih},
  {Bodensteiner}, {Crowther}, {Gieles}, {Gromadzki}, {H{\'e}nault-Brunet},
  {Herrero}, {de Koter}, {Iwanek}, {Koz{\l}owski}, {Lennon}, {Ma{\'\i}z
  Apell{\'a}niz}, {Mr{\'o}z}, {Moffat}, {Picco}, {Pietrukowicz}, {Poleski},
  {Rybicki}, {Schneider}, {Skowron}, {Skowron}, {Soszy{\'n}ski},
  {Szyma{\'n}ski}, {Toonen}, {Udalski}, {Ulaczyk}, {Vink}, \&
  {Wrona}}]{Shenar2022}
{Shenar}, T., {Sana}, H., {Mahy}, L., {et~al.} 2022, Nature Astronomy, 6, 1085

\bibitem[{{Sim{\'o}n-D{\'\i}az} \& {Herrero}(2007)}]{Simon-Diaz2007}
{Sim{\'o}n-D{\'\i}az}, S. \& {Herrero}, A. 2007, \aap, 468, 1063

\bibitem[{{Sim{\'o}n-D{\'\i}az} \& {Herrero}(2014)}]{Simon-Diaz2014}
{Sim{\'o}n-D{\'\i}az}, S. \& {Herrero}, A. 2014, \aap, 562, A135

\bibitem[{{Smee} {et~al.}(2013){Smee}, {Gunn}, {Uomoto}, {Roe}, {Schlegel},
  {Rockosi}, {Carr}, {Leger}, {Dawson}, {Olmstead}, {Brinkmann}, {Owen},
  {Barkhouser}, {Honscheid}, {Harding}, {Long}, {Lupton}, {Loomis}, {Anderson},
  {Annis}, {Bernardi}, {Bhardwaj}, {Bizyaev}, {Bolton}, {Brewington}, {Briggs},
  {Burles}, {Burns}, {Castander}, {Connolly}, {Davenport}, {Ebelke}, {Epps},
  {Feldman}, {Friedman}, {Frieman}, {Heckman}, {Hull}, {Knapp}, {Lawrence},
  {Loveday}, {Mannery}, {Malanushenko}, {Malanushenko}, {Merrelli}, {Muna},
  {Newman}, {Nichol}, {Oravetz}, {Pan}, {Pope}, {Ricketts}, {Shelden},
  {Sandford}, {Siegmund}, {Simmons}, {Smith}, {Snedden}, {Schneider},
  {SubbaRao}, {Tremonti}, {Waddell}, \& {York}}]{Smee2013}
{Smee}, S.~A., {Gunn}, J.~E., {Uomoto}, A., {et~al.} 2013, \aj, 146, 32

\bibitem[{{Smith}(1976)}]{Smith1976b}
{Smith}, M.~A. 1976, \apj, 203, 603

\bibitem[{{Smith} \& {Gray}(1976)}]{Smith1976}
{Smith}, M.~A. \& {Gray}, D.~F. 1976, \pasp, 88, 809

\bibitem[{{Tonry} \& {Davis}(1979)}]{Tonry1979}
{Tonry}, J. \& {Davis}, M. 1979, \aj, 84, 1511

\bibitem[{Wilks(1938)}]{Wilks1938}
Wilks, S.~S. 1938, The Annals of Mathematical Statistics, 9, 60

\bibitem[{{Willcox} {et~al.}(2025){Willcox}, {Marchant}, {Vigna-G{\'o}mez},
  {Sana}, {Bodensteiner}, {Deshmukh}, {Esseldeurs}, {Fabry},
  {H{\'e}nault-Brunet}, {Janssens}, {Mahy}, {Patrick}, {Pauli}, {Renzo},
  {Sander}, {Shenar}, {van Son}, \& {Stoop}}]{Willcox2025}
{Willcox}, R., {Marchant}, P., {Vigna-G{\'o}mez}, A., {et~al.} 2025, \aap, 700,
  A59

\bibitem[{{Zucker} \& {Mazeh}(1994)}]{Zucker1994}
{Zucker}, S. \& {Mazeh}, T. 1994, \apj, 420, 806

\end{thebibliography}

\begin{appendix}
\nolinenumbers



\section{Functional form of the kernels} \label{app: kernels}
The Lorentzian and Gaussian components exponentially attenuate the signal in Fourier space, namely
\begin{equation}
\begin{aligned}
\tilde{K}_{\gamma}(\zeta)
\propto \,\, 
\exp\left[-{\pi v_\gamma}|\zeta|\right] \quad \text{and} \quad 
\tilde{K}_{\textsc{g}}(\zeta)
\propto \,\, 
\exp\left[-\frac{\pi^2 v_{\textsc{g}}^2}{4\ln2}\zeta^2\right].
\end{aligned}
\end{equation}
The rotational kernel is given by the expression
\begin{equation}\label{eq: rotational kernel}
\begin{aligned}
\tilde{K}_{\textsc{r}}(\zeta)
 \propto \quad 
\frac{6(1-\epsilon)}{x(3-\epsilon)} J_1(x) \quad + 
\frac{6\epsilon}{x^3(3-\epsilon)}\left(\sin x - x\cos x\right),
\end{aligned}
\end{equation}
where $J_1$ is the first-order Bessel function and $x$ is a dimensionless variable defined as $x \equiv 2\pi \zeta v_{\textsc{r}}$. The Fourier transform of the rotational kernel with linear limb darkening is outlined by \citet[][see chapter 18 therein]{Gray1992}; we reproduce it here in our velocity-coordinate convention. 

\section{Linearized noise propagation}\label{app: noise model}

Equations~(\ref{eq: single line model}) and~(\ref{eq: phase model}) are written with an additive noise term, although the fitted observables are nonlinear functions of the measured spectrum. We make this approximation explicit here, since it also defines the weights used in the following appendix sections. The Fourier transform is linear, so measurement noise added to equation~(\ref{eq: sb1 convolution}) remains additive after transformation. For analytic propagation, we assume independent Gaussian pixel noise with constant variance $\sigma_{\rm pix}^2$, neglecting its variation across the absorption line. The real and imaginary parts of the transformed noise then have a common standard deviation, $\sigma$, whose value depends on the window and Fourier-transform convention. For observed spectra, we calibrate the noise scale and amplitude-search null distribution empirically, without modeling the full Fourier covariance.

However, taking the logarithm of the transformed spectra is not a linear operation. As long as the noise is subdominant, it can be expanded to first order. The logarithm of the Fourier-transformed observed spectrum is therefore
\begin{equation}\label{eq: complex log expansion}
\ln \tilde{S}_{\rm obs}
\simeq
\ln \tilde{S}
+
\frac{\tilde{\varepsilon}}{\tilde{S}},
\end{equation}
where $\tilde{\varepsilon}$ is the Fourier transformed noise. 
The two observables used in this work are the real and imaginary parts of this
expression: the log-amplitude of equation~(\ref{eq: single line model}) and the
phase of equation~(\ref{eq: phase model}). In this limit $\tilde{\varepsilon}$ is circularly symmetric, and the two parts are therefore independent, zero-mean, and Gaussian, with the same standard deviation. For the $i^{\text{th}}$ conjugate wavelength, 
\begin{equation}\label{eq: log noise scaling}
\sigma_i \simeq \frac{\sigma}{|\tilde{S}_i|}.
\end{equation}
In the analysis of observed spectra, $\sigma$ is instead calibrated empirically. Synthetic single-star profiles are injected into line-free regions of the same spectra and refitted, and $\sigma$ is rescaled so that the median reduced $\chi^2$ of these fits equals unity (Section~\ref{sec: worked example}). This is approximately the theoretical median for a correctly calibrated noise model. The rescaling also accounts for low-frequency structure not described by $\sigma_{\rm pix}$.

The expansion in equation~(\ref{eq: complex log expansion}) assumes that the Fourier amplitude is large compared with the noise. Its accuracy degrades gradually as the signal-to-noise ratio decreases, with appreciable deviations appearing below ${\sim}\,3$ in our simulations. In this regime, the log-amplitude becomes biased by the noise floor and the likelihood can develop secondary minima. A related effect arises when evaluating equation~(\ref{eq: log noise scaling}). The noise-free Fourier amplitude is not known and is replaced in practice by the observed one. Since the observed Fourier amplitude is the absolute value of a noisy complex quantity, small amplitudes tend to be biased towards larger values. Substituting it into equation~(\ref{eq: log noise scaling}) therefore tends to underestimate the per-mode uncertainty, particularly when the amplitude approaches the noise scale. The same noise realization also enters both the log-amplitude residuals and their weights, making the formal parameter uncertainties optimistic.

We account for these effects through injection--recovery experiments that repeat the complete analysis over independent noise realizations. Fitting the injected single-star profiles with the same procedure as the data yields (i) the empirical noise scale $\sigma$ described above, (ii) the null distribution of the detection statistic (Section~\ref{sec: detection}), and (iii) the sampling distribution of the fitted parameters, from which the formal uncertainty is rescaled so that its nominal 68\% interval has the correct coverage about the median of the recovered values. In all cases the fitted range is defined by a fixed limit in conjugate wavelength rather than by a threshold on the measured amplitudes. The limit is placed where the noise-free model amplitude approaches the noise level. Modes near the limit therefore lie at signal-to-noise ratios of order unity to three, at the edge of the linearized regime, and are retained rather than removed by a per-mode cut because they carry much of the line-shape information; the resulting optimism of the formal uncertainties is the effect that step (iii) calibrates. For the simulated example of Figure~\ref{fig: line example} the noise scale is known by construction and only step (iii) is applied; for the observed spectra of Section~\ref{sec: worked example} we apply steps (i) and (ii), and the parameter uncertainties quoted there are formal.

 \section{Expected precision} \label{app: expected precision}
\subsection{Rotation estimation for a single line} \label{app: rotation normal equations}
For a given rotational velocity, the model is linear in the remaining parameters and can be written in matrix form 
\begin{equation}
\mathbf{y} - \mathbf{k} = M\,\mathbf{a}  + \boldsymbol{\varepsilon},
\end{equation}
where $M$ is the design matrix, $[\mathbf{1}, ~ \boldsymbol{\zeta} ,~\boldsymbol{\zeta}^2]$, and $\mathbf{k}$ is a vector representing the rotational kernel's contribution at each $\zeta$ for a fixed rotational velocity value. The model parameters are estimated via standard weighted least squares with a diagonal inverse-variance weight matrix, as per equation~\eqref{eq: log noise scaling},
\begin{equation}
W_{ij} = \delta_{ij}\,\sigma_i^{-2}.
\end{equation}
For a given $v_{\textsc{r}}$ and $\epsilon$, the inferred parameters are
\begin{equation}
\hat{\mathbf{a}}= (M^\top W M)^{-1}~M^\top W~(\mathbf{y}-\mathbf{k}).
\end{equation}

Given these parameters, we can calculate the profiled likelihood, evaluated along the direction orthogonal to the subspace spanned by the linear terms. The profiled residuals, which retain only the dependence on the rotational velocity, are then
\begin{equation}
\mathbf{e} \equiv (I-P)\,(\mathbf{y}-\mathbf{k}),
\end{equation}
where 
\begin{equation*}
P \equiv M\,(M^\top W M)^{-1} M^\top W
\end{equation*}
is a projection operator that can be calculated in advance. The corresponding weighted objective function then becomes
\begin{equation}
\chi^2 = \mathbf{e}^\top\, W\,\mathbf{e}.
\end{equation}
By minimizing the objective function, one can infer the rotational velocity of the star.

The sensitivity of the model to the projected rotational velocity depends only on the part of the rotational kernel that is not absorbed by the remaining polynomial terms. This implies that the Fisher information is
\begin{equation}
\mathcal{I}_{v_\textsc{r}} \approx
\left(\frac{d\mathbf{k}}{dv_\textsc{r}}\right)^\top (I-P)^\top W\,(I-P)\,\frac{d\mathbf{k}}{dv_\textsc{r}}.
\end{equation}
Only the part of the rotational kernel that is not absorbed by the polynomial baseline contributes to this expression. In the limit of slow rotation, $k$ can be expanded as in equation~(\ref{eq: lnS approx}), such that
\begin{equation*}
\left|\frac{d k}{dv_\textsc{r}}\right|
\propto 
v_\textsc{r}^{3}\,\zeta^{4} + \text{lower-order terms}.
\end{equation*}
The terms up to $\zeta^2$ are projected out by $P$ and therefore do not contribute information.

Substituting this scaling into the Fisher information yields the Cram\'er--Rao lower bound on the fractional uncertainty,
\begin{equation}\label{eq: vr crlb}
\frac{\Delta v_\textsc{r}}{v_\textsc{r}}
\;\gtrsim\;
\left(\frac{6}{c}\right)\frac{\sigma}{\alpha_{\textsc{a}}} 
\end{equation}
where 
\begin{equation*}
c^2 \equiv {\sum_i |\tilde{S}_i|^2\alpha_{\textsc{a}}^{-2}\,(\pi \zeta_i v_\textsc{r})^8 }.
\end{equation*}
If the Fourier amplitudes decay rapidly, such that the signal reaches the noise floor already at $\zeta_i \ll v_\textsc{r}^{-1}$, the weighted sum becomes small. In this regime the available rotational information is strongly reduced and the fractional uncertainty in $v_\textsc{r}$ becomes large. 

Here we retain only the leading term of the rotational contribution. This approximation assumes that the finer structure of the rotational kernel lies below the noise level. With sufficiently high signal-to-noise ratio, however, additional structure becomes measurable. In particular, the first zero of the Bessel function in the rotational kernel may be resolved, which would further improve the precision. In that regime the polynomial expansion of the kernel is no longer valid, and the full rotational profile must be considered. In many practical cases, however, the rotational velocity is only a few times the instrumental line-spread function width, or the signal-to-noise ratio is limited, so the leading-order scaling derived above provides an adequate approximation. This implies that, if rotational broadening dominates the line width (${\rm FWHM}\sim v_\textsc{r}$), the fractional error scales as
\begin{equation}\label{eq: vr fractional error scaling}
\frac{\Delta v_\textsc{r}}{v_\textsc{r}}
\propto
\alpha_{\textsc{a}}^{-1}\,\sigma. 
\end{equation}
 A similar approach can be used to determine the expected fractional error in the estimation of the Voigt profile parameters.

\subsection{Sensitivity for companion detection}

We now estimate the attainable precision in the line-depth ratio, $\alpha$. For a fixed velocity separation $v$, the model can be written as
\begin{equation}
\mathbf{y} = M\,\mathbf{a} + \alpha\,\mathbf{t} + \boldsymbol{\varepsilon},
\end{equation}
where ${t}_i=\cos(2\pi \zeta_i v)$ is the binary modulation template. The design matrix and measurement vector are as defined above.

Profiling over the linear baseline removes any contribution that can be represented by the fitted polynomial terms.  
The inverse-variance weight matrix $W$ and projection operator $P$ are defined as in the previous section.  
The remaining sensitivity to the line-depth ratio is therefore carried by the projected modulation, $(I-P)\,\mathbf{t}$. The corresponding Fisher information is
\begin{equation}
\mathcal{I}_\alpha = \mathbf{t}^\top (I-P)^\top \,W\,(I-P)\,\mathbf{t}.
\end{equation}
This expression isolates the measurable component of the binary modulation, excluding any slowly varying structure absorbed by the baseline.  

In the limit of small velocity separation, the cosine term approaches unity and becomes degenerate with the polynomial baseline.  
As a result, when both stars share the same velocity, the line-depth ratio becomes unmeasurable.  
After projection, $(I-P)\,\mathbf{t}\approx \frac{2}{3} \pi^4\zeta^4 v^4$, since the constant and quadratic terms are absorbed by the fitted baseline.  

As before, substituting this scaling into the Fisher information yields an approximate lower bound on the fractional uncertainty,
\begin{equation}\label{eq: alpha crlb}
\frac{\Delta \alpha}{\alpha}
\;\gtrsim\;
\bigg(\frac{v_{\textsc{g}}}{v}\bigg)^4
\left(\frac{3}{2c}\right)\frac{\sigma}{\alpha_{\textsc{b}}} 
\end{equation}
where
\begin{equation*}
c^2 \equiv {\sum_i |\tilde{S}_i|^2\alpha_{\textsc{a}}^{-2}\,(\pi \zeta_i v_{\textsc{g}})^8 }.
\end{equation*}
If the Fourier amplitudes decay rapidly, such that the signal reaches the noise floor already at $\zeta_i \ll v^{-1}$, the weighted sum becomes small. In this regime the available companion information is strongly reduced and the fractional uncertainty in $\alpha$ becomes large.

\section{Validity of the LSF-limit approximation}\label{app: approx validity}

The cosine model in equation~(\ref{eq: unresolved binary model}) is the limiting case in which the two components have the same effective kernel over the fitted Fourier range, so that $\mathcal{R}(\zeta)\simeq 1$. When the instrumental LSF dominates both line profiles this assumption is natural. If the intrinsic or rotational broadening differs between the stars, the kernel ratio must be retained.

If the secondary line is broader than the primary, $\tpker{b}$ decays more rapidly than $\tpker{a}$. In this case $\mathcal{R}(\zeta)$ is usually smooth over the useful range of $\zeta$, before the Fourier amplitude reaches the noise floor. The single scalar coefficient multiplying the cosine can then be replaced by a slowly varying function, for example a low-order polynomial in $\zeta$. This gives a simple extension of equation~(\ref{eq: unresolved binary model}), valid in the small-$\alpha$ regime as long as the detailed structure of the secondary kernel is unresolved. The opposite case is less benign. If the primary line is much broader than the secondary, $\mathcal{R}(\zeta)$ can grow rapidly with $\zeta$. Then the companion perturbation is no longer described by a
cosine with a slowly varying amplitude, and both the constant-$\mathcal{R}$ model and its polynomial extension can fail. Such systems should instead be modeled with the full Fourier log-amplitude in equation~(\ref{eq: lnS explicit}), or with direct spectral modeling of equation~(\ref{eq: binary_system}), analogous to \texttt{TODCOR}
\citep{Zucker1994}.

\begin{figure*}
    \centering
    \includegraphics[width=1\linewidth]{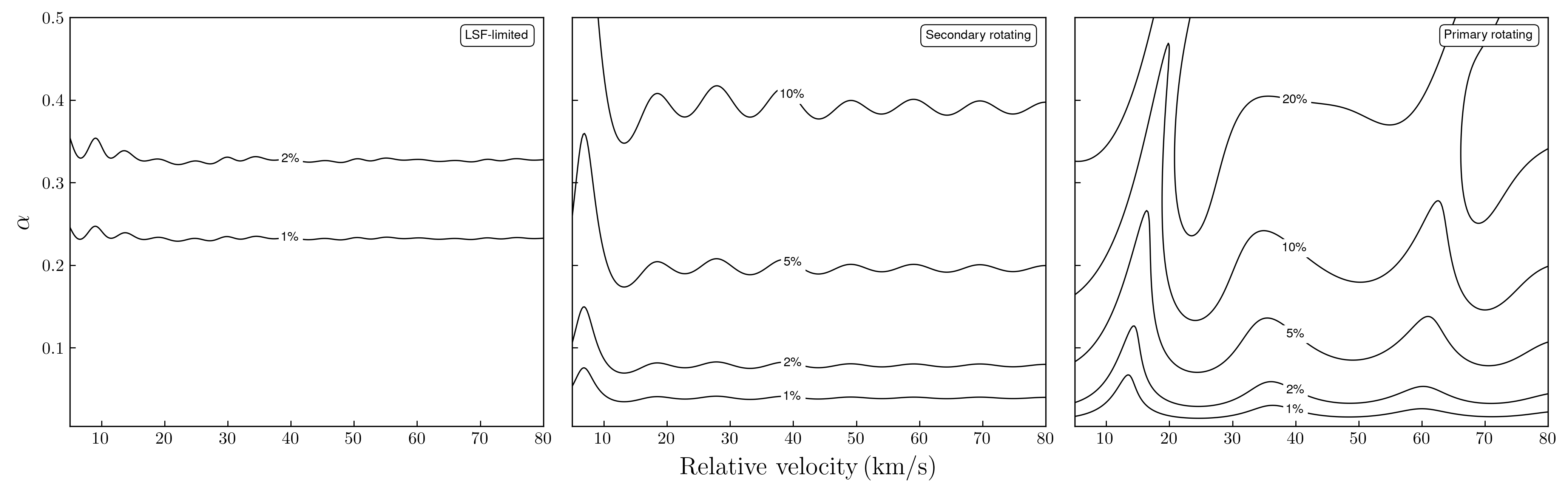}
\caption{
Accuracy of the single-cosine approximation for the binary correction to the Fourier log-amplitude. Contours show the root-mean-square modeling error, normalized by the root-mean-square amplitude of the exact binary correction, in the $(v,\alpha)$ plane. The mismatch is computed on a Fourier-amplitude grid truncated at $\zeta_{\max}=1/{\rm FWHM}$, where FWHM is the quadrature sum of the primary's Gaussian, Lorentzian, and rotational widths ($\zeta_{\max}\simeq0.05~\mathrm{s\,km^{-1}}$ in the left and middle panels and $\simeq0.02~\mathrm{s\,km^{-1}}$ in the right panel). In the left panel, both components are LSF-limited. They have identical profiles, with $v_{\textsc{g}}=20\,\mathrm{km\,s^{-1}}$, $v_\gamma=5\,\mathrm{km\,s^{-1}}$, and no rotation, so the kernel ratio term is unity. In the middle panel, the primary is unchanged and the secondary is rotationally broadened by $20\,\mathrm{km\,s^{-1}}$, in addition to the fiducial LSF. As a result, the secondary kernel decays more rapidly in Fourier space. In the right panel, the primary is also rotationally broadened by $20\,\mathrm{km\,s^{-1}}$, while the secondary is LSF-limited. This is the least favorable case for the approximation, because the primary kernel is narrower in Fourier space and the kernel ratio can grow rapidly. 
}
    \label{fig: approx accuracy}
\end{figure*}

Figure~\ref{fig: approx accuracy} illustrates these regimes. The left panel shows the fractional modeling error of $\ln |\tilde{S}|$ in the limiting case where the line-spread function dominates the profiles of both stars. As expected, the approximation improves toward smaller $\alpha$, since equation~(\ref{eq: unresolved binary model}) retains only the leading term in the companion contribution. The middle panel shows a case where the primary remains effectively unresolved but the secondary is rapidly rotating. The approximation becomes less accurate because $\mathcal{R}(\zeta)$ is no longer close to unity, but the error remains limited because the secondary kernel decays quickly in Fourier space. The right panel shows the most challenging case for the simplified model, where the primary is rapidly rotating. In this regime, $\mathcal{R}(\zeta)$ grows rapidly and the approximation breaks down at smaller velocity separations and lower line-depth ratios. These systems are not necessarily difficult to detect, since a narrow secondary superposed on a broad primary is often visible in wavelength space, but they require the more general modeling described above.

\end{appendix}
\end{document}